\documentclass[sigconf]{acmart}
\setcopyright{acmcopyright}
\usepackage{booktabs} 
\usepackage{microtype}
\usepackage[latin1]{inputenc}
\usepackage[OT2,T1]{fontenc}
\usepackage{amssymb}
\usepackage[linesnumbered,ruled,vlined]{algorithm2e}
\usepackage{algpseudocode}
\usepackage{bm}
\usepackage{amsmath}
\usepackage{graphicx}
\usepackage{epsfig}
\usepackage{url}
\usepackage{psfrag}
\usepackage{balance}
\usepackage{subfigure}
\usepackage{float}
\usepackage{multirow}
\usepackage{pstricks}
\usepackage{array}
\usepackage{enumitem}
\usepackage{hyperref}
\usepackage{pifont}
\graphicspath{{figs/}}

\def\BibTeX{{\rm B\kern-.05em{\sc i\kern-.025em b}\kern-.08emT\kern-.1667em\lower.7ex\hbox{E}\kern-.125emX}}

\newtheorem{theorem}{\textbf{Theorem}}
\newenvironment{pf}{{\it \textbf{Proof.} }}{\hfill $\square$\medskip}

\copyrightyear{2019} 
\acmYear{2019} 
\setcopyright{acmcopyright}
\acmConference[KDD '19]{The 25th ACM SIGKDD Conference on Knowledge Discovery and Data Mining}{August 4--8, 2019}{Anchorage, AK, USA}
\acmBooktitle{The 25th ACM SIGKDD Conference on Knowledge Discovery and Data Mining (KDD '19), August 4--8, 2019, Anchorage, AK, USA}
\acmPrice{15.00}
\acmDOI{10.1145/3292500.3330825}
\acmISBN{978-1-4503-6201-6/19/08}

\usepackage{balance}

\begin{document}

\title{A Memory-Efficient Sketch Method for Estimating\\High Similarities in Streaming Sets}

\author{
     Pinghui Wang$^{1,2,\star}$, Yiyan Qi$^{1,\star}$, Yuanming Zhang$^{1}$, Qiaozhu Zhai$^{1}$, Chenxu Wang$^{2,*}$, 
}
\author{John C.S. Lui$^{3}$, Xiaohong Guan$^{2,1,4,*}$}
\thanks{$^\star$Pinghui Wang and Yiyan Qi contributed equally to this work.}
\thanks{$^*$Corresponding Author.}
\affiliation{%
	\institution{$^{1}$NSKEYLAB, Xi'an Jiaotong University, Xi'an, China}
	\institution{$^{2}$Shenzhen Research Institute of Xi'an Jiaotong University, Shenzhen, China}
	\institution{$^{3}$The Chinese University of Hong Kong, Hong Kong}
	\institution{$^{4}$Department of Automation and NLIST Lab, Tsinghua University, Beijing, China}
}
\email{{phwang,qzzhai,cxwang,xhguan}@mail.xjtu.edu.cn, {qiyiyan,zhangyuanming}@stu.xjtu.edu.cn,} \email{cslui@cse.cuhk.edu.hk}

\renewcommand{\shortauthors}{Pinghui Wang, et al.}

\begin{abstract}
	Estimating set similarity and detecting highly similar sets are fundamental problems in areas such as databases, machine learning, and information retrieval.
	MinHash is a well-known technique for approximating Jaccard similarity of sets and has been successfully used for many applications such as similarity search and large scale learning.
	Its two compressed versions, $b$-bit MinHash and Odd Sketch, can significantly reduce the memory usage of the original MinHash method, especially for estimating high similarities (i.e., similarities around 1).
	Although MinHash can be applied to static sets as well as streaming sets, of which elements are given in a streaming fashion and cardinality is unknown or even infinite,
	unfortunately, $b$-bit MinHash and Odd Sketch fail to deal with streaming data.
	To solve this problem,
	we design a memory efficient sketch method, \emph{MaxLogHash}, to accurately estimate Jaccard similarities in streaming sets.
	Compared to MinHash, our method uses smaller sized registers (each register consists of less than 7 bits) to build a compact sketch for each set.
	We also provide a simple yet accurate estimator for inferring Jaccard similarity from MaxLogHash sketches.
	In addition, we derive formulas for bounding the estimation error and determine the smallest necessary memory usage (i.e., the number of registers used for a MaxLogHash sketch) for the desired accuracy.
	We conduct experiments on a variety of datasets,
	and experimental results show that our method MaxLogHash is about 5 times more memory efficient than MinHash with the same accuracy and computational cost for estimating high similarities.
\end{abstract}

\ccsdesc[500]{Mathematics of computing~Probabilistic algorithms}
\ccsdesc[500]{Information systems~Similarity measures}
\ccsdesc[500]{Theory of computation~Sketching and sampling}

%
\keywords{Streaming algorithms;Sketch;Jaccard coefficient similarity}

\maketitle

\section{Introduction} \label{sec:introduction}
Data streams are ubiquitous in nature.
Examples range from financial transactions to Internet of things (IoT) data, network traffic, call logs,
trajectory logs, etc.
Due to the nature of these applications which involve massive volume of data, 
it is prohibitive to collect the entire data streams, especially when computational and storage resources are limited \cite{Li2018Approximate}.
Therefore, it is important to develop memory efficient methods such as sampling and sketching techniques for mining large streaming data.

Many datasets can be viewed as collections of sets
and computing set similarities is fundamental for a variety of applications in areas such as databases, machine learning, and information retrieval.
For example, one can view each mobile device's trajectory as a set and each element in the set corresponds to a tuple of time $t$ and the physical location of the device at time $t$.
Then, mining devices with similar trajectories is useful for identifying friends or devices belonging to the same person.  
Other examples are datasets encountered in computer networks, mobile phone networks, and online social networks (OSNs),
where learning user similarities in the sets of users' visited websites on the Internet,
connected phone numbers, and friends on OSNs is fundamental for applications such as link prediction and friendship recommendation.

One of the most popular set similarity measures is the Jaccard similarity coefficient,
which is defined as $\frac{|A\cap B|}{|A\cup B|}$ for two sets $A$ and $B$.
To handle large sets,
MinHash (or, minwise hashing)~\cite{Broder2000} is a powerful set similarity estimation technique,
which uses an array of $k$ registers to build a sketch for each set.
Its accuracy only depends on the value of $k$ and the Jaccard similarity of two sets of interest,
and it is independent from the size of two sets.
MinHash has been successfully used for a variety of applications,
such as similarity search~\cite{BroderSEQUENCES1997},
compressing social networks~\cite{ChierichettiKDD2009},
advertising diversification~\cite{GollapudiWWW2009},
large scale learning~\cite{LiNIPS2011},
and web spam detection~\cite{UrvoyTOW2008}.
Many of these applications focus on estimating similarity values close to 1.  
Take similar document search in a sufficiently large corpus as an example.
For a corpus,
there may be thousands of documents which are similar to the query document,
therefore our goal is not just to find similar documents, 
but also to provide a short list (e.g., top-10) and ranking of the most
similar documents. 
For such an application, we need effective methods that are very accurate and memory-efficient for estimating high similarities.
To achieve this goal, there are two compressed MinHash methods, $b$-bit MinHash~\cite{PingWWW2010} and Odd Sketch~\cite{MitzenmacherWWW14}, which were proposed in the past few years to further reduce the memory usage of the original MinHash by dozens of times, while to provide comparable estimation accuracy especially for large similarity values.
However, we observe that these two methods fail to handle data streams (the details will be given in Section~\ref{sec:preliminaries}).

To solve the above challenge,
recently, Yu and Weber~\cite{YuArxiv2017} develop a method, HyperMinHash.
HyperMinHash consists of $k$ registers,
whereas each register has two parts, an FM (Flajolet-Martin) sketch~\cite{Flajolet1985} and a $b$-bit string.
The $b$-bit string is computed based on the fingerprints (i.e., hash values) of set elements that are mapped to the register.
Based on HyperMinHash sketches of two sets $A$ and $B$,
HyperMinhash first estimates $|A\cup B|$  and then infers the Jaccard similarity of $A$ and $B$ from the number of collisions of $b$-bit strings
given $|A\cup B|$.
Later in our experiments, we demonstrate that HyperMinHash not only exhibits a large bias for high similarities, but it is also computationally expensive for estimating similarities,
which results in a large estimation error and a big delay in querying highly similar sets.
More importantly, it is difficult to analytically analyze the estimation bias and variance of HyperMinHash,
which are of great value in practice--the bias and variance can be used to bound an estimate' error and determine the smallest necessary sampling budget (i.e., $k$) for
a desired accuracy.
In this paper, we develop a novel memory efficient method, \emph{MaxLogHash}, to estimate Jaccard similarities in streaming sets.
Similar to MinHash, MaxLogHash uses a list of $k$ registers to build a compact sketch for each set.
Unlike MinHash which uses a 64-bit (resp. 32-bit) register for storing the minimum hash value of 64-bit (resp. 32-bit) set elements,
our method MaxLogHash uses only 7-bit register (resp. 6-bit register) to approximately record the logarithm value of the minimum hash value,
and this results in 9 times (resp. 5 times) reduction in memory usage.
Another attractive property is that our MaxLogHash sketch can be computed incrementally,
therefore, MaxLogHash is able to handle streaming-sets.
Given any two sets' MaxLogHash sketches, we provide a simple yet accurate estimator for their Jaccard similarity, and derive exact formulas for bounding the estimation error.
We conduct experiments on a variety of publicly available datasets,
and experimental results show that our method MaxLogHash reduces the amount of memory required for MinHash by 5 folds to achieve the same desired accuracy and computational cost.

The rest of this paper is organized as follows.
The problem formulation is presented in Section~\ref{sec:problem}.
Section~\ref{sec:preliminaries} introduces preliminaries used in this paper.
Section~\ref{sec:method} presents our method MaxLogHash.
The performance evaluation and testing results are presented in Section~\ref{sec:results}.
Section~\ref{sec:related} summarizes related work. Concluding remarks then follow.

\section{Problem Formulation} \label{sec:problem}
For ease of reading and comprehension,
we say that each set belongs to a user,
elements in the set are items (e.g., products) that the user connects to.
Let $U$ denote the set of users and $I$ denote the set of all items.
Let $\Pi=e^{(1)} e^{(2)} \cdots e^{(t)} \cdots$ denote the user-item stream of interest,
where $e^{(t)} = (u^{(t)}, i^{(t)})$ is the element of $\Pi$ occurred at discrete time $t> 0$,
$u^{(t)}\in U$ and $i^{(t)}\in I$ are the element's user and item,
which represents a connection from user $u^{(t)}$ to item $i^{(t)}$.
We assume that $\Pi$ has no duplicate user-item pairs\footnote{Duplicated user-item pairs can be easily checked and filtered using fast and memory-efficient techniques such as Bloom filter~\cite{BloomACMCommun1970}.},
that is, $e^{(i)}\ne e^{(j)}$ when $i\ne j$.
Let $I_u^{(t)}\subset I$ be the item set of user $u\in U$,
which consists of items that user $u$ connects to before and including time $t$.
Let $\cup^{(t)}(u_1, u_2)$ denote the union of two sets $I_{u_1}^{(t)}$ and $I_{u_2}^{(t)}$,
that is,
$
\cup^{(t)}(u_1, u_2) = I_{u_1}^{(t)} \cup I_{u_2}^{(t)}.
$
Similarly, we define the intersection of two sets $I_{u_1}^{(t)}$ and $I_{u_1}^{(t)}$ as
$
\cap^{(t)}(u_1, u_2) = I_{u_1}^{(t)} \cap I_{u_2}^{(t)}.
$
Then, the Jaccard similarity of sets $I_{u_1}^{(t)}$ and $I_{u_2}^{(t)}$ is defined as
\[
J_{u_1, u_2}^{(t)}=\frac{|\cap^{(t)}(u_1, u_2)|}{|\cup^{(t)}(u_1, u_2)|},
\]
which reflects the similarity between users $u_1$ and $u_2$.
In this paper, we aim to develop a fast and accurate method to estimate $J_{u_1, u_2}^{(t)}$ for any two users $u_1$ and $u_2$ over time,
and to detect pairs of high similar users.
When no confusion arises, we omit the superscript $(t)$ to ease exposition.


\section{Preliminaries}\label{sec:preliminaries}
In this section, we first introduce MinHash~\cite{Broder2000}.
Then, we elaborate two state-of-the-art memory-efficient methods \emph{$b$-bit MinHash}~\cite{PingWWW2010} and \emph{Odd Sketch}~\cite{MitzenmacherWWW14} that can decrease the memory usage of the original MinHash method.
At last, we demonstrate that both $b$-bit MinHash and Odd Sketch fail to handle streaming sets.

\subsection{MinHash} \label{sec:minhash}
Given a random permutation (or hash function\footnote{MinHash assumes no hash collisions.}) $\pi$ from elements in $I$ to elements in $I$, i.e., a hash function maps integers in $I$ to \emph{distinct} integers in $I$ at random.
Broder et al.~\cite{Broder2000} observed that the Jaccard similarity of two sets $A, B\subseteq I$ equals
\[
J_{A,B} = \frac{|\cap(A,B)|}{|\cup(A,B)|} = P(\min(\pi (A)) = \min(\pi (B))),
\]
where $\pi (A) = \{\pi(w): w\in A\}$.
Therefore, MinHash uses a sequence of $k$ independent permutations $\pi_1, \ldots, \pi_k$
and estimates $J_{A,B}$ as
\[
\hat J_{A,B} = \frac{\sum_{i=1}^k \mathbf{1}(\min(\pi_1 (A))  = \min(\pi_1 (B))}{k},
\]
where $\mathbf{1}(\mathbb{P})$ is an indicator function that equals 1 when the predicate $\mathbb{P}$ is true and 0 otherwise.
Note that $\hat J_{A,B}$ is an unbiased estimator for $J_{A, B}$, i.e., $\mathbb{E}(\hat J_{A,B}) = J_{A,B}$,
and its variance is
\[
\text{Var}(\hat J_{A,B}) = \frac{J_{A,B}(1 - J_{A,B})}{k}.
\]
Therefore, instead of storing a set $A$ in memory, one can compute and store its MinHash sketch $S_A$,
i.e.,
\[
S_A = (\min(\pi_1 (A)), \ldots, \min(\pi_k (A))),
\]
which reduces the memory usage when $|A|> k$.
The Jaccard similarity of any two sets can be accurately and efficiently estimated based on their MinHash sketches.
\subsection{b-bit MinHash} \label{sec:bbit}
Li and K{\"{o}}nig~\cite{PingWWW2010} proposed a method, $b$-bit MinHash,  to further reduce the memory usage.
$b$-bit MinHash reduces the memory required for storing a MinHash sketch $S_A$  from $32 k$ or $64k$ bits\footnote{A 32- or 64-bit register is used to store each $\min(\pi_i(A))$, $i=1,\ldots, k$.} to $bk$ bits.
The basic idea behind $b$-bit MinHash is that the same hash values give the same lowest $b$ bits
while two different hash values give the same lowest $b$ bits with a small probability $1/2^b$.
Formally, let $\min^{(b)}(\pi(A))$ denote the lowest $b$ bits of the value of $\min(\pi(A))$ for a permutation $\pi$.
Define the $b$-bit MinHash sketch of set $A$ as
\[
S_A^{(b)} = (\text{min}^{(b)}(\pi_1 (A)), \ldots, \text{min}^{(b)}(\pi_k (A))).
\]
To mine set similarities,
Li and K{\"{o}}nig~\cite{PingWWW2010} first compute $S_A$ for each set $A$,
and then store its $b$-bit MinHash sketch $S_A^{(b)}$.
At last, the Jaccard similarity $J_{A, B}$ is estimated as
\[
\hat J_{A,B}^{(b)} = \frac{\sum_{i=1}^k \mathbf{1}(\text{min}^{(b)}(\pi_i (A))  = \text{min}^{(b)}(\pi_i (B)))- k/2^b}{k(1 - 1/2^b)}.
\]
$\hat J_{A,B}^{(b)}$ is also an unbiased estimator for $J_{A, B}$,
and its variance is
\[
\text{Var}(\hat J_{A,B}^{(b)}) = \frac{1 - J_{A,B}}{k}\left(J_{A,B} + \frac{1}{2^b-1}\right).
\]

\subsection{Odd Sketch}
Mitzenmacher et al.~\cite{MitzenmacherWWW14}
developed a method Odd Sketch,
which is more memory efficient than $b$-bit MinHash when mining sets of high similarity.
Odd Sketch uses a hash function $h$ that maps each tuple $(i, \min(\pi_i (A)))$, $i=1, \ldots, k$,
to an integer in $\{1,\ldots, z\}$ at random.
For a set $A$,
its odd sketch $S_A^\text{(Odd)}$ consists of $z$ bits.
Function $h$ maps tuples $(1, \min(\pi_1 (A))), \ldots, (k, \min(\pi_k (A)))$ into $z$ bits of $S_A^\text{(Odd)}$ at random.
$S_A^\text{(Odd)}[j]$, $1\le j\le z$, is the parity of the number of tuples that are mapped to the $j^\text{th}$ bit of $S_A^\text{(Odd)}$.
Formally, $S_A^\text{(Odd)}[j]$ is computed as
\[
S_A^\text{(Odd)}[j] = \oplus_{i=1,\ldots,k} \mathbf{1}(h(i, \min(\pi_i (A))) = j), \quad 1\le j\le z.
\]
The Jaccard similarity $J_{A,B}$ is then estimated as
\[
\hat J_{A,B}^\text{(Odd)} = 1 + \frac{z}{4k} \ln \left(1 - \frac{2\sum_{i=1}^z S_A^\text{(Odd)}[j] \oplus S_B^\text{(Odd)}[j]}{z} \right).
\]
Mitzenmacher et al. demonstrate that $\hat J_{A,B}^\text{(Odd)}$ is more accurate than $\hat J_{A,B}^\text{(b)}$
under the same memory usage (refer to~\cite{MitzenmacherWWW14} for details of the error analysis of $\hat J_{A,B}^\text{(Odd)}$).

\subsection{Discussion}
\noindent\textbf{MinHash can be directly applied to stream data.} We can easily find that
MinHash sketch can be computed incrementally.
That is, one can compute the MinHash sketch of set $A\cup \{v\}$ from the MinHash sketch of set $A$ as
\[
\min(\pi (A\cup \{v\})) = \min(\min(\pi (A)), \pi(v)).
\]
\noindent\textbf{Variants $b$-bit MinHash and Odd Sketch cannot be used to handle streaming sets.}
Let $\pi^{(b)}(v)$ denote the lowest $b$ bits of $\pi(v)$.
Then, one can easily show that
\[
\text{min}^{(b)}(\pi (A\cup \{v\})) \ne \text{min}(\text{min}^{(b)}(\pi (A)), \pi^{(b)}(v)).
\]
It shows that computing $\text{min}^{(b)}(\pi (A\cup \{v\}))$ requires the hash value $\pi(w)$ of each $w\in A\cup \{v\}$.
In addition, we observe that $\min^{(b)}(\pi(A))$ cannot be approximated as $\min_{w\in A}\pi^{(b)}(w)$, which can be computed incrementally,
because $\min_{w\in A}\pi^{(b)}(w)$ equals 0 with a high probability when $|A|\gg 2^b$.
Similarly, we cannot compute the odd sketch of a set incrementally.
Therefore, both $b$-bit MinHash and Odd Sketch fail to deal with streaming sets.

\section{Our Method}\label{sec:method}
\subsection{Basic Idea}
Let $h$ be a function that maps any element $v$ in $I$ to a random number in range $(0,1)$.
i.e., $h(v)\sim Uniform(0, 1)$.
Define the log-rank of $v$ with respect to hash function $h$ as
$
r(v) \gets \lfloor -  \log_2 h (v) \rfloor.
$
We compute and store
\[
\text{MaxLog}(h(A)) = \max_{v\in A} r(v) =\max_{v\in A} \lfloor -  \log_2 h (v) \rfloor.
\]
Let us now develop a simple yet accurate method to estimate Jaccard similarity of streaming sets based on the following properties of function $\text{MaxLog}(h(A))$.\\
\noindent\textbf{Observation 1.}  $\text{MaxLog}(h(A))$ can be represented by an integer of no more than $\lceil\log_2\log_2|I|\rceil$ bits with a high probability.
For each $v\in I$, we have $h(v)\sim Uniform(0, 1)$, and thus
$
r(v)\sim Geometric(1/2),
$
supported on the set $\{0, 1, 2, \ldots\}$,
that is,
\[
P(r(v) = j) = \frac{1}{2^{j+1}}, \quad P(r(v) < j) = 1-\frac{1}{2^j}, \quad j\in \{0, 1, 2, \ldots\}.
\]
Then, one can easily find that
\begin{equation*}
P(\text{MaxLog}(h(A)) \le 2^{\lceil\log_2\log_2|I|\rceil}-1) = \left(1 - \frac{1}{2^{2^{\lceil\log_2\log_2|I|\rceil}}}\right)^{|A|}.
\end{equation*}
For example, when $A\subseteq \{1,\ldots, 2^{64}\}$ and $|A|\le 2^{54}$, we only require 6 bits to store $\text{MaxLog}(h(A))$
with probability at least 0.999.

\noindent\textbf{Observation 2.}  $\text{MaxLog}(h(A))$ can be computed incrementally. This is because
\[
\text{MaxLog}(h(A\cup \{v\}))=\max(\text{MaxLog}(h(A)), \lfloor -  \log_2 h (v) \rfloor).
\]

\noindent\textbf{Observation 3.} $J_{A,B}$ can be easily estimated from $\text{MaxLog}(h(A))$ and $\text{MaxLog}(h(B))$
with a little additional information.
We find that
\begin{equation*}
\begin{split}
\gamma=&P(\text{MaxLog}(h(A))\ne \text{MaxLog}(h(B)))\\
=&  \sum_{j=1}^{+\infty} \frac{|A\setminus B|}{2^{j+1}}\left(1-\frac{1}{2^{j+1}}\right)^{|A\setminus B|-1} \left(1-\frac{1}{2^j}\right)^{|B|}\\
&+\sum_{j=1}^{+\infty} \frac{|B\setminus A|}{2^{j+1}}\left(1-\frac{1}{2^{j+1}}\right)^{|B\setminus A|-1} \left(1-\frac{1}{2^j}\right)^{|A|}.
\end{split}
\end{equation*}
Due to the limited space, we omit the details of how $\gamma$ is derived.
Similar to MinHash, we have $P(\max(h(A))\ne \max(h(B))) = 1-J_{A, B}$.
Therefore, we have $\gamma< 1-J_{A, B}$.
Although $\gamma$ can be estimated similar to MinHash using $k$ hash functions $h_1, \ldots, h_k$,
that is,
\[
\mathbb{E}(\gamma) = \frac{\sum_{i=1}^k \mathbf{1}(\text{MaxLog}(h_i(A))\ne \text{MaxLog}(h_i(B)))}{k},
\]
unfortunately, it is difficult to compute $J_{A,B}$ from $\gamma$.
To solve this problem, we observe
\begin{equation*}
P(\text{MaxLog}(h(A))\ne \text{MaxLog}(h(B)) \wedge \delta_{A,B}=1)\approx 0.7213 (1-J_{A,B}),
\end{equation*}
where $\delta_{A,B}=1$ indicates that there exists one and only one element in $A\cup B$ of which log-rank equals $\text{MaxLog}(h(A\cup B))$.

Based on the above three observations, we propose to incrementally and accurately estimate the value of $P(\text{MaxLog}(h(A))\ne \text{MaxLog}(h(B)) \wedge \delta_{A,B}=1)$ using $k$ hash functions $h_1, \ldots, h_k$.
Then, we easily infer the value of $J_{A,B}$.

\subsection{Data Structure}
The MaxLogHash sketch of a user $u$, i.e., $S_u$, consists of $k$ bit-strings,
where each bit-string $S_u[i], 1\le i\le k$, has two components, $s_u[i]$ and $m_u[i]$, i.e.,
$
S_u[i] = s_u[i] \parallel m_u[i].
$
At any time $t$,
$m_u[i]$ records the maximum hash value of items in $I_u^{(t)}$ with respect to hash function $r_i(\cdot)=\lfloor -  \log_2 h_i (\cdot) \rfloor$,  i.e., $m_u[i] = \max_{w\in I_u^{(t)}} r_i(w)$,
where $I_u^{(t)}$ refers to the set of items that user $u$ connected to before and including time $t$;
$s_u[i]$ consists of 1 bit and its value indicates whether there exists one and only one item $w\in I_u$ such that $r_i(w) = m_u[i]$.
As we mentioned, we can use $\lceil\log_2\log_2|I|\rceil$ bits to record the value of $m_u[i]$ with a high probability (very close to 1).
When $m_u[i]\ge 2^{\lceil\log_2\log_2|I|\rceil}$, we use a hash table to record tuples $(u, i, m_u[i])$  for all users.

\subsection{Update Procedure}
For each user $u\in U$,
when it first connects with an item $w$ in stream $\Pi$,
we initialize the MaxLogHash sketch of user $u$ as
$
S_u[i] = 1\parallel r_i(w), \quad i=1, \ldots, k,
$
where $r_i(w) = \lfloor -  \log_2 h_i (w) \rfloor$.
That is, we set indicator $s_u[i] = 1$ and register $m_u[i] = r_i(w)$.
For any other item $v$ that user $u$ connects to after the first item $w$,
i.e., an user-item pair $(u,v)$ occurring on stream $\Pi$ after the user-item pair $(u,w)$,
we update it as follows:
We first compute the log-rank of item $v$, i.e., $r_i(v) = \lfloor -  \log_2 h_i (v) \rfloor$, $i=1, \ldots, k$.
When $r_i(v)$ is smaller than $m_u[i]$, we perform no further operations for updating the user-item $(u,v)$.
When $r_i(v)=m_u[i]$, it indicates that at least two items in $I_u$ has a log-rank value $m_u[i]$.
Therefore, we simply set $s_u[i]=0$.
When $r_i(v)>m_u[i]$, we set
$S_u[i] = 1\parallel r_i(v)$.

\subsection{Jaccard Similarity Estimation}
Define variables
\[
\chi_{u_1, u_2} [i] = \mathbf{1}(m_{u_1}[i] \ne m_{u_2}[i]), \quad i = 1, \ldots, k,
\]
\[
\psi_{u_1, u_2} [i] =\left\{
                    \begin{array}{ll}
                      s_{u_1}[i], & m_{u_1}[i] > m_{u_2}[i] \\
                      s_{u_2}[i], & m_{u_1}[i] < m_{u_2}[i]\\
                      -1, & m_{u_1}[i] = m_{u_2}[i].
                    \end{array}
                  \right.
\]
Let $\delta_{u_1, u_2} [i] = \mathbf{1}(\chi_{u_1, u_2} [i] = 1) \mathbf{1}(\psi_{u_1, u_2} [i] = 1)$.
Note that $\delta_{u_1, u_2}[i]=1$ indicates that there exists one and only one element in set $\cup(u_1, u_2)$ of which log-rank equals $\max_{w\in \cup(u_1, u_2)} r_i(w)$
with respect to function $r_i$.
Then, we have the following theorem.
\begin{theorem}\label{theorem: prob}
For non-empty sets $I_{u_1}$ and $I_{u_2}$,
we have $P(\delta_{u_1, u_2} [i] = 1) = 0$, $i=1, \ldots, k$, when $|\cup(u_1, u_2)|=1$.
Otherwise, we have
\begin{equation*}
P(\delta_{u_1, u_2} [i] = 1)= \alpha_{|\cup(u_1, u_2)|} (1-J_{u_1, u_2}), \quad i = 1, \ldots, k,
\end{equation*}
where 
$
\alpha_n = n \sum_{j=1}^{+\infty} \frac{1}{2^{j+1}}\left(1-\frac{1}{2^j}\right)^{n-1},  \quad n \ge 2.
$
\end{theorem}

\begin{pf}
	Let $r^*$ be the maximum log-rank of all items in $\cup (u_1, u_2)$.
	When two items $w$ and $v$ in $I_{u_1}$ or $I_{u_2}$ has the log-rank value $r^*$,
	we easily find that $\psi_{u_1, u_2} [i]=0$.
	When only one item $w$ in $I_{u_1}$ and only one item $v$ in $I_{u_2}$ have the log-rank value $r^*$,
	we easily find that $\chi_{u_1, u_2} [i]=0$.
	Let
	\[
	\Delta(u_1, u_2) = (I_{u_1}\setminus I_{u_2}) \cup (I_{u_2}\setminus I_{u_1}) =\cup (u_1, u_2) \setminus \cap(u_1, u_2).
	\]
	Then, we find that event $\chi_{u_1, u_2} [i] = 1 \wedge \psi_{u_1, u_2} [i] = 1$ happens (i.e., $\delta_{u_1, u_2} [i]=1$) only when one item $w$ in $\Delta(u_1, u_2)$ has a log-rank value larger than all items in $\cup (u_1, u_2)\setminus \{w\}$.
	For any item $v\in I$, we have $h_i(v)\sim Uniform(0, 1)$ and so $r_i(v)\sim Geometric(1/2)$,
	supported on the set $\{0, 1, 2, \ldots\}$.
	Based on the above observations, when $|\cup(u_1, u_2)|\ge 2$, we have
	\begin{equation*}
	\begin{split}
	&P(\delta_{u_1, u_2} [i]=1 \wedge r^* = j) \\
	&= \sum_{w\in \Delta(u_1, u_2)} P(r_i(w) = j) \prod_{v\in \cup (u_1, u_2)\setminus \{w\}}  P(r_i(v) < j)\\
	&= \frac{|\Delta(u_1, u_2)|}{2^{j+1}}\left(1-\frac{1}{2^j}\right)^{|\cup(u_1, u_2)|-1}.
	\end{split}
	\end{equation*}
	Therefore, we have
	\begin{equation*}
	\begin{split}
	&P(\delta_{u_1, u_2} [i]=1)=\sum_{j=0}^{+\infty} P(\delta_i = 1\wedge r^* = j) \\
	&= \sum_{w\in \Delta(u_1, u_2)} P(r_w = j) \prod_{v\in \cup (u_1, u_2)\setminus \{w\}}  P(r_v < j)\\
	&= \sum_{j=0}^{+\infty} \frac{|\Delta(u_1, u_2)|}{2^{j+1}}\left(1-\frac{1}{2^j}\right)^{|\cup(u_1, u_2)|-1}\\
	&= \sum_{j=1}^{+\infty} \frac{|\Delta(u_1, u_2)|}{|\cup(u_1, u_2)|} \cdot \frac{|\cup(u_1, u_2)|}{2^{j+1}}\left(1-\frac{1}{2^j}\right)^{|\cup(u_1, u_2)|-1}\\
	&=\alpha_{|\cup(u_1, u_2)|} (1-J_{u_1, u_2}),
	\end{split}
	\end{equation*}
	where the last equation holds because $|\Delta(u_1, u_2)| = |\cup(u_1, u_2)| - |\cap(u_1, u_2)|$.
\end{pf}

\begin{figure}[t]
	\center
	\includegraphics[width=0.35\textwidth]{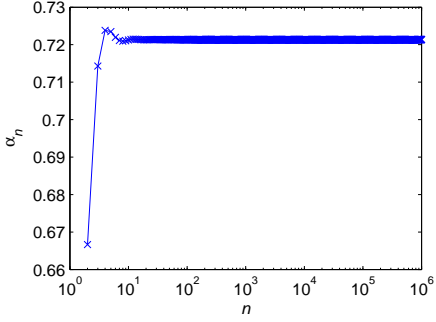}
	\caption{Value of $\alpha_n$, $n=2,\ldots, 10^6$ where $|\alpha_2 - 0.7213| = 0.0546$, $|\alpha_n - 0.7213| \le 0.007$ when $n\ge 3$.}
	\label{fig:alphan}
\end{figure}

Define variable
$
\hat k = \sum_{i=1}^k \mathbf{1} (\delta_{u_1, u_2} [i]=1).
$
From Theorem~\ref{theorem: prob}, the expectation of $\hat k$ is computed as
\begin{eqnarray}\label{eq:expk}
\mathbb{E}(\hat k) &=& \mathbb{E}\left(\sum_{i=1}^k \mathbf{1} (\delta_{u_1, u_2} [i]=1)\right) \nonumber\\
 &=& \sum_{i=1}^k \mathbb{E}(\mathbf{1} (\delta_{u_1, u_2} [i]=1)) \nonumber\\
 &=& k \alpha_{|\cup(u_1, u_2)|} (1-J_{u_1, u_2}).
\end{eqnarray}
Therefore, we have
\[
J_{u_1, u_2} = 1 - \frac{\mathbb{E}(\hat k)}{k\alpha_{|\cup(u_1, u_2)|}}.
\]
Note that the cardinality of set $\cup(u_1, u_2)$ (i.e. $|\cup(u_1, u_2)|$) is unknown.
To solve this challenge,
we find that
\begin{equation*}
\begin{split}
\alpha_n &= \frac{n}{2} \sum_{j=1}^{+\infty} \frac{1}{2^j}\left(1-\frac{1}{2^j}\right)^{n-1}\\
&= \frac{n}{2} \sum_{j=1}^{+\infty} \frac{1}{2^j} \sum_{l=0}^{n-1} \binom{n-1}{l} \left(-\frac{1}{2^j}\right)^{n-l-1}\\
&= \frac{n}{2}  \sum_{l=0}^{n-1} (-1)^{n-l-1} \binom{n-1}{l}  \sum_{j=1}^{+\infty} \frac{1}{2^{j(n-l)}}\\
&= \frac{n}{2}  \sum_{l=0}^{n-1} (-1)^{n-l-1} \binom{n-1}{l}  \frac{1}{2^{n-l}-1}.
\end{split}
\end{equation*}
Figure~\ref{fig:alphan} shows that the value of $\alpha_n$, $n=2,3,\ldots.$
We easily find that $\alpha_n\approx \alpha=0.7213$ when $n\ge 2$.
Therefore, we estimate $J_{u_1, u_2}$ as
\[
\hat J_{u_1, u_2} = 1 - \frac{\hat k}{k\alpha}.
\]


\subsection{Error Analysis}\label{sec:error}
The error of our method MaxLogHash is shown in the following theorem.
\begin{theorem}\label{theorem: error}
For any users $u_1, u_2\in U$, we have
\[
\mathbb{E}(\hat J_{u_1, u_2}) - J_{u_1, u_2} = (1 - \beta_{|\cup(u_1, u_2)|})(1 - J_{u_1, u_2}),
\]
where $\beta_n = \frac{\alpha_n}{\alpha}$.
The variance of $\hat J_{u_1, u_2}$ is computed as
\begin{equation*}
\text{Var}(\hat J_{u_1, u_2})=\frac{\beta_{|\cup(u_1, u_2)|} (1-J_{u_1, u_2})(\alpha^{-1} - \beta_{|\cup(u_1, u_2)|} (1-J_{u_1, u_2}))}{k}.
\end{equation*}
When $|\cup(u_1, u_2)|\ge 3$, 
we have $|\beta_{|\cup(u_1, u_2)|}-1|\le 0.01$,
and so $\mathbb{E}(\hat J_{u_1, u_2}) \approx J_{u_1, u_2}$ and $\text{Var}(\hat J_{u_1, u_2})\approx \frac{(1-J_{u_1, u_2})(J_{u_1, u_2}+0.3864)}{k}$.
\end{theorem}

\begin{pf}
	From equation~(\ref{eq:expk}), we  easily have
	\begin{equation*}
	\begin{split}
	\mathbb{E}(\hat J_{u_1, u_2}) &=\mathbb{E}\left(1 - \frac{\hat k}{k\alpha}\right)\\
	&=1 - \frac{k \alpha_{|\cup(u_1, u_2)|} (1-J_{u_1, u_2})}{k \alpha}\\
	&=\beta_{|\cup(u_1, u_2)|} J_{u_1, u_2} + 1 - \beta_{|\cup(u_1, u_2)|}.
	\end{split}
	\end{equation*}
	To derive $\text{Var}(\hat J_{u_1, u_2})$, we first compute
	\begin{eqnarray*}
		\mathbb{E}(\hat k^2) &=& \mathbb{E}\left(\left(\sum_{i=1}^k \mathbf{1} (\delta_{u_1, u_2} [i]=1)\right)^2\right) \\
		&=&\sum_{i=1}^k \mathbb{E}\left((\mathbf{1} (\delta_{u_1, u_2} [i]=1))^2 \right) \\
		&&+\sum_{i\ne j, 1\le i,j\le k} \mathbb{E}\left(\mathbf{1} (\delta_{u_1, u_2} [i]=1) \mathbf{1} (\delta_{u_1, u_2} [j]=1) \right) \\
		&=& k \alpha_{|\cup(u_1, u_2)|} (1-J_{u_1, u_2}) + k (k-1) \alpha_{|\cup(u_1, u_2)|}^2 (1-J_{u_1, u_2})^2.
	\end{eqnarray*}
	Then, we have
	\begin{eqnarray}\label{eq:expk2}
	&&\text{Var}(\hat k) = \mathbb{E}(\hat k^2) - (\mathbb{E}(\hat k))^2\nonumber\\
	&&= k \alpha_{|\cup(u_1, u_2)|} (1-J_{u_1, u_2})(1 - \alpha_{|\cup(u_1, u_2)|} (1-J_{u_1, u_2})).
	\end{eqnarray}
	From the definition of $\hat J_{u_1, u_2}$, we have
	\begin{equation*}
	\text{Var}(\hat J_{u_1, u_2}) = \text{Var}\left(1 - \frac{\hat k}{k\alpha}\right) = \frac{\text{Var}(\hat k)}{k^2 \alpha^2}.
	\end{equation*}
	Then, we easily obtain a closed-form formals of $\text{Var}(\hat J_{u_1, u_2})$ from equation~(\ref{eq:expk2}).
\end{pf}

\subsection{Reduce Processing Complexity}\label{subsec:extension}
Inspired by OPH (one permutation hashing)~\cite{Linips2012},
which significantly reduces the time complexity of MinHash for processing each element in the set,
we can use a hash function which splits items in $I_u$ into $k$ registers at random,
and each register $S_u[i]$, $1\le i\le k$, records $\text{MaxLog}(h(\{v: v\in I_u \wedge h(v) = j\}))$
as well as the value of indicator $s_u[i]$,
which is similar to the regular MaxLogHash method.
We name this extension as MaxLogOPH.
MaxLogOPH reduces the time complexity of processing each item from $O(k)$ to $O(1)$.
When $|u_1\cup u_2|\gg k$, our experiments demonstrate that MaxLogOPH is comparable to MaxLogHash in terms of accuracy.
\section{Evaluation} \label{sec:results}
The algorithms are implemented in Python, and run on a computer with a Quad-Core Intel(R) Xeon(R) CPU E3-1226 v3 CPU 3.30GHz processor.
To demonstrate the reproducibility of the experimental results,
we make our source code publicly available\footnote{http://nskeylab.xjtu.edu.cn/dataset/phwang/code/MaxLog.zip}.
\begin{figure*}[t]
	\centering
	\subfigure[(balanced) RMSE, $k=128$]{
		\includegraphics[width=0.22\textwidth]{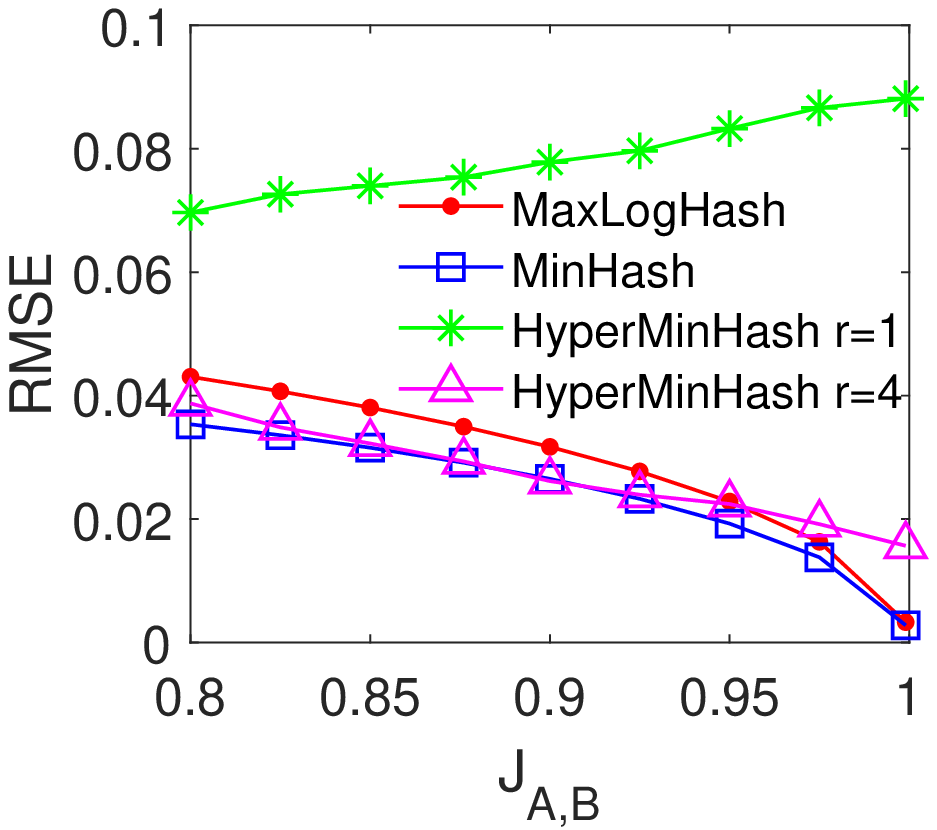}}
	\subfigure[(balanced) RMSE, $k=256$]{
		\includegraphics[width=0.22\textwidth]{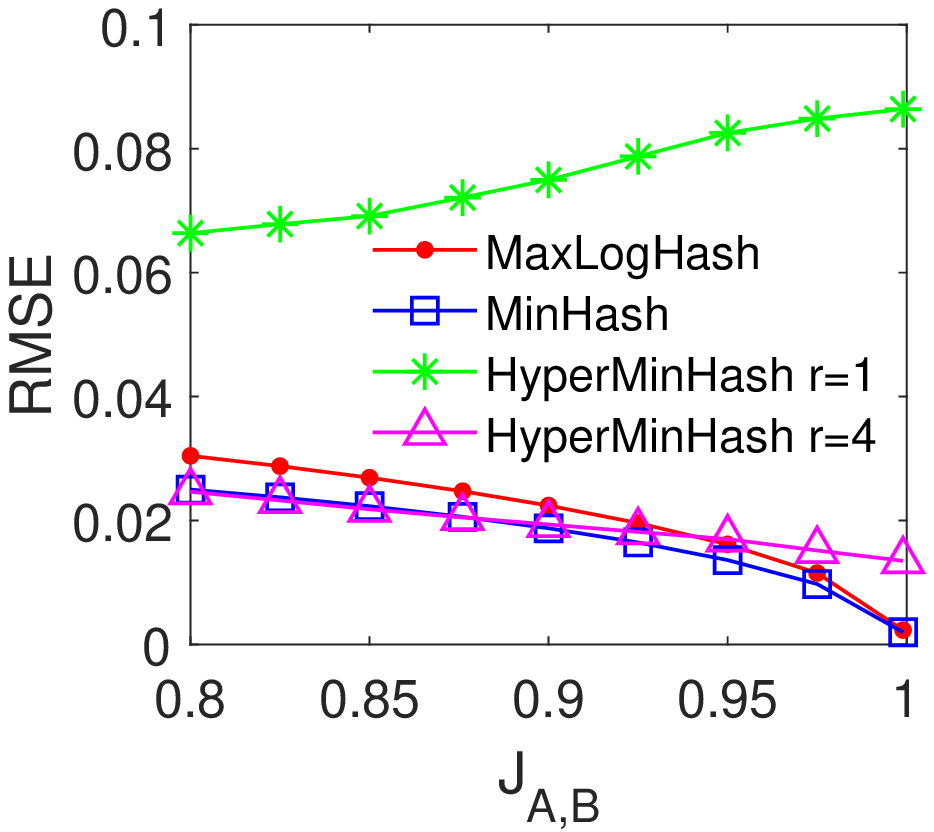}}
	\subfigure[(unbalanced) RMSE, $k=128$]{
		\includegraphics[width=0.22\textwidth]{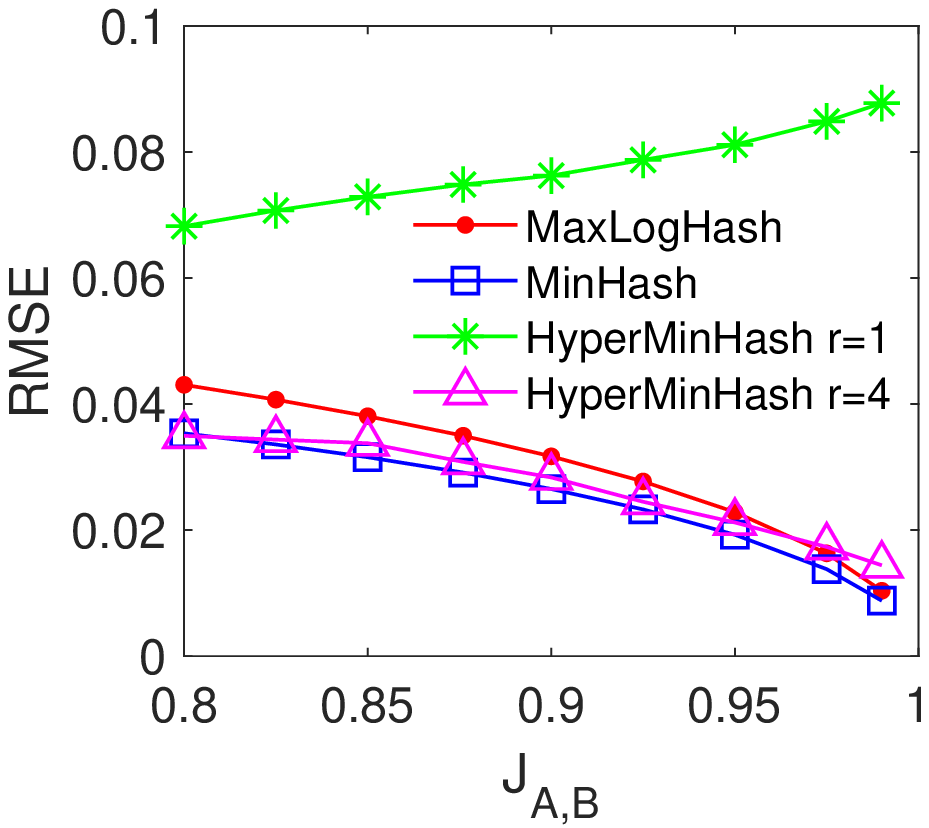}}
	\subfigure[(unbalanced) RMSE, $k=256$]{
		\includegraphics[width=0.22\textwidth]{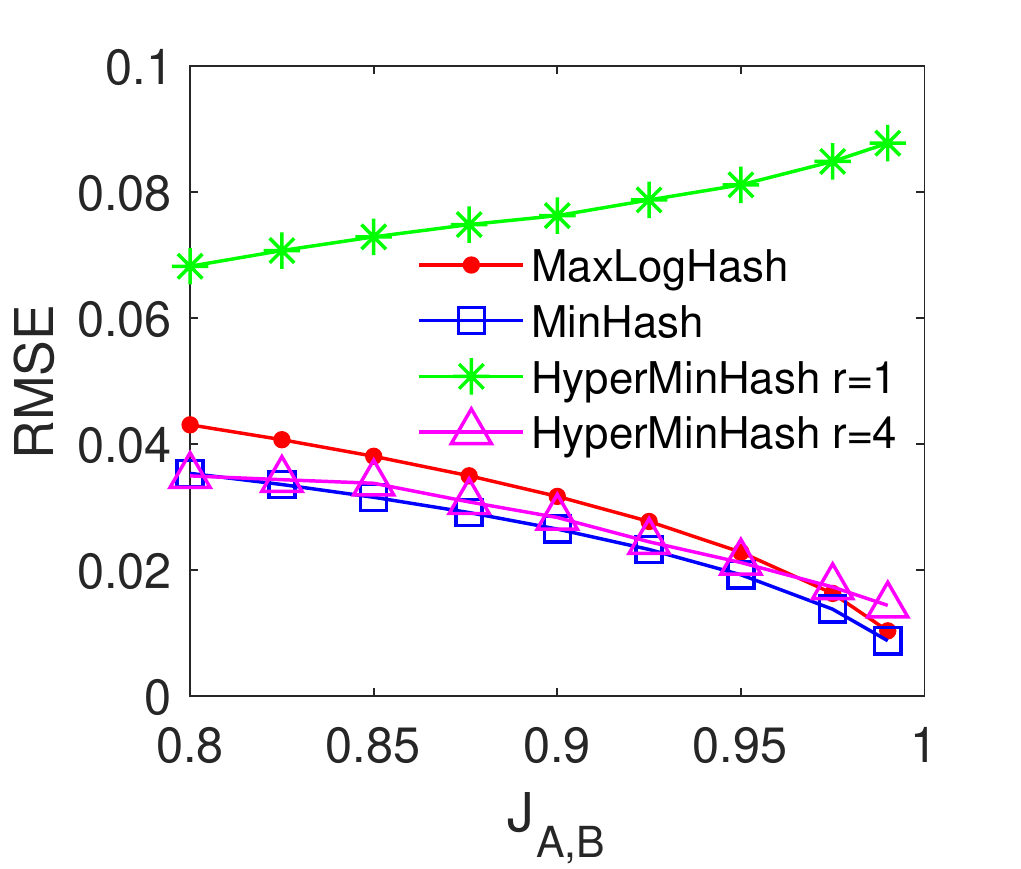}}
	\subfigure[(balanced) Bias, $k=128$]{
		\includegraphics[width=0.22\textwidth]{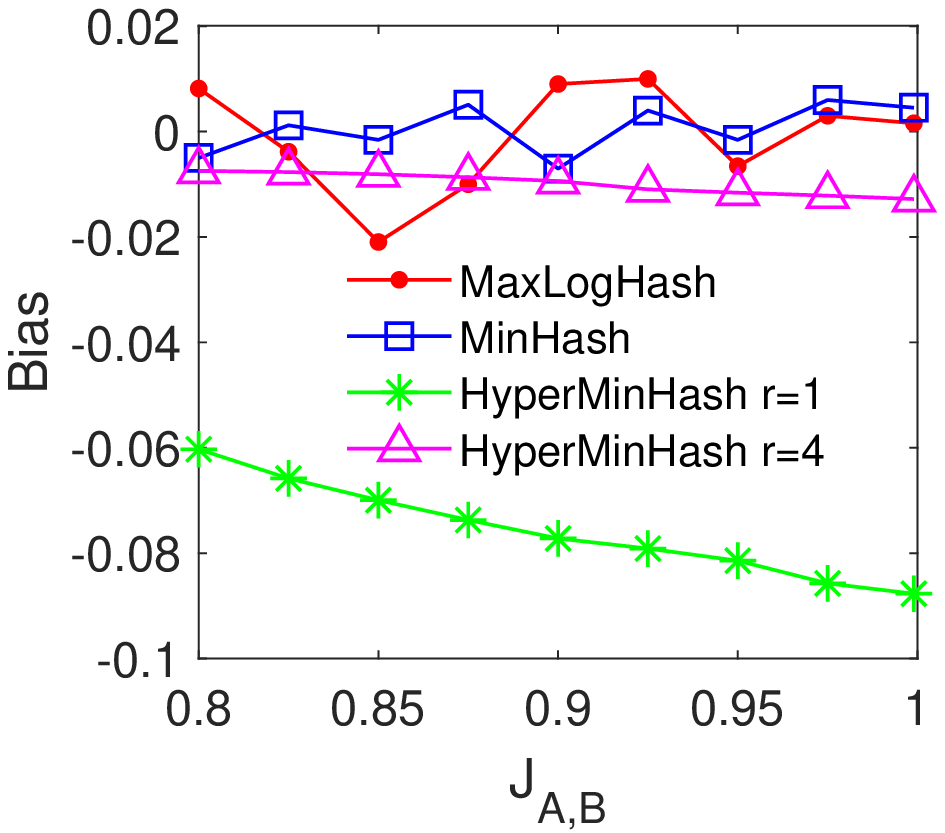}}
	\subfigure[(balanced) Bias, $k=256$]{
		\includegraphics[width=0.22\textwidth]{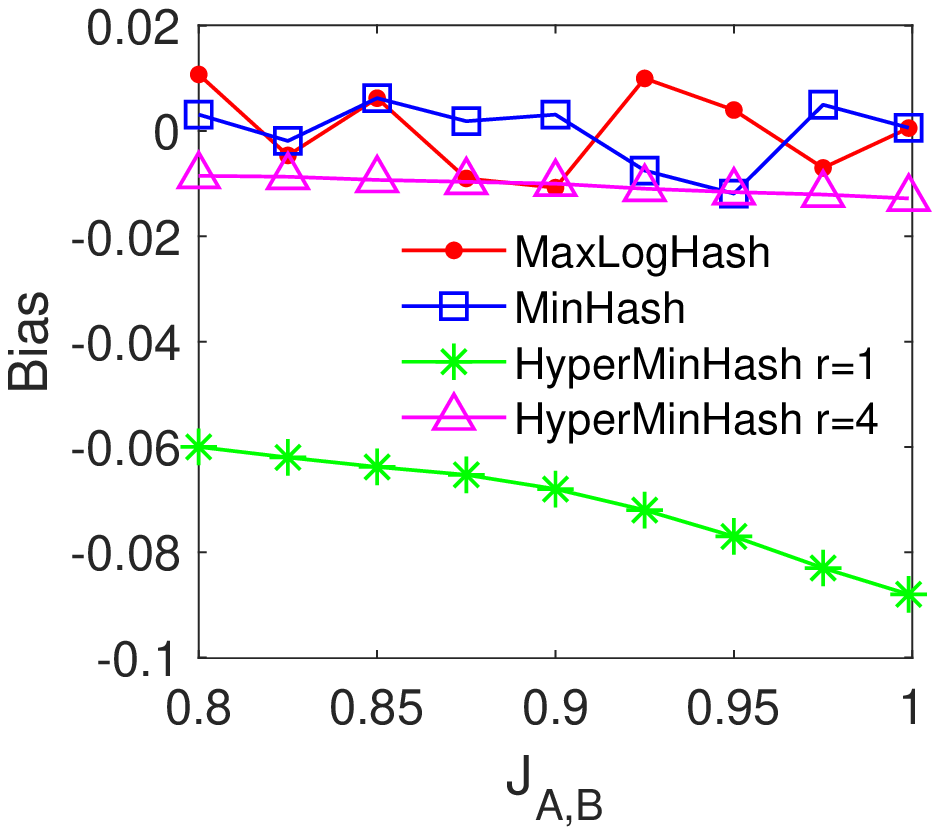}}
	\subfigure[(unbalanced) Bias, $k=128$]{
		\includegraphics[width=0.22\textwidth]{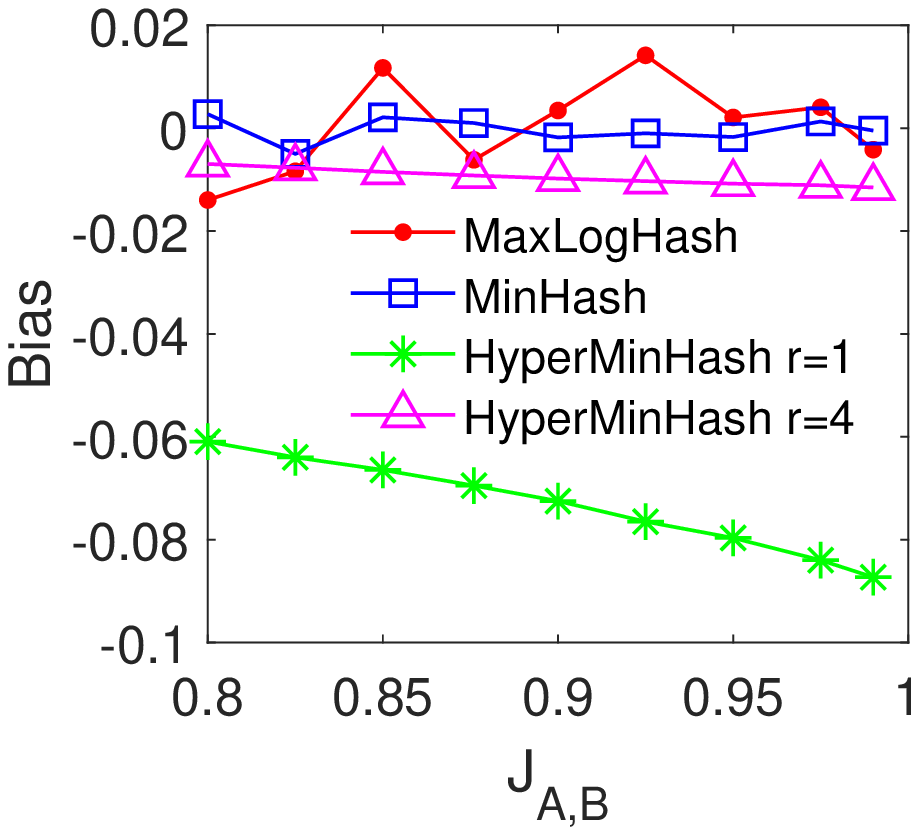}}
	\subfigure[(unbalanced) Bias, $k=256$]{
		\includegraphics[width=0.22\textwidth]{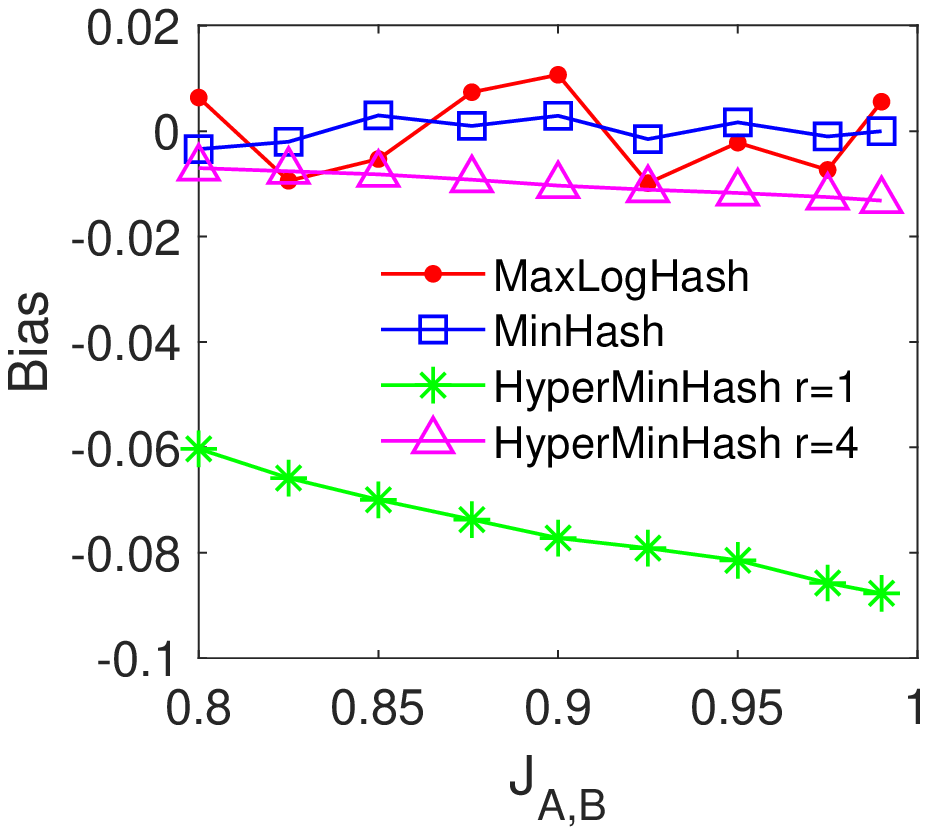}}
	\caption{Estimation error of our method MaxLogHash in comparison with MinHash and HyperMinHash on both balanced and unbalanced set-pairs.}\label{fig:synerror}
\end{figure*}

\subsection{Datasets}
For simplicity, we assume that elements in sets are 32-bit numbers,
i.e., $I=\{0,1,\ldots,2^{32}-1\}$.
We evaluate the performance of our method MaxLogHash a variety of datasets.

1) \textbf{Synthetic datasets.} 
Our synthetic datasets consist of set-pairs $A$ and $B$ with various cardinalities and Jaccard similarities. We conduct our experiments on the following two different settings:

\noindent $\bullet$ \textbf{Balanced set-pairs} (i.e., $|A|=|B|$). We set $|A|=|B|=n$ and vary $J_{A,B}$ in $\{0.80,0.81,...,1.00\}$.
Specially, we generate set $A$ by randomly selecting $n$ different numbers from $I$
and generate set $B$ by randomly selecting $|A \cap B|=\frac{J_{A,B}|A|}{1+J_{A,B}}$ different numbers from set $A$ and $n - |A \cap B|$ different numbers from set $I\setminus A$.
In our experiments, we set $n=10,000$ by default.

\noindent $\bullet$ \textbf{Unbalanced set-pairs} (i.e., $|A|\ne|B|$). We set $|A| = n$ and $|B|= J_{A,B} n$, where we vary $J_{A,B} \in \{0.80,0.81,...,0.99\}$.
Specially, we generate set $A$ by randomly selecting $n$ different numbers from $I$
and generate set $B$ by selecting $J_{A,B} n$ different elements from $A$.

2) \textbf{Real-world datasets.}
Similar to~\cite{MitzenmacherWWW14},
we evaluate the performance of our method on the detection of item-pairs (e.g., pairs of products) that always appear together in the same records (e.g., transactions).
We conduct experiments 
on two real-world datasets\footnote{http://fimi.ua.ac.be/data/}: MUSHROOM and CONNECT,
which are also used in~\cite{MitzenmacherWWW14}.
We generate a stream of item-record pairs for each dataset,
where a record can be viewed as a transaction and items in the same record can be viewed as products bought together.
For each record $x$ in the dataset of interest and every item $w$ in $x$, we append an element $(w, x)$ to the stream of item-record pairs.
In summary, MUSHROOM and CONNECT have $8,124$ and $67,557$ records,
$119$ and $127$ distinct items, and $186,852$ and $2,904,951$ item-record pairs, respectively.

\subsection{Baselines}
Our methods use $k$ $6$-bit registers to build a sketch for each set.
We compare our methods with the following state-of-the-art methods:
\noindent $\bullet$ \textbf{MinHash~\cite{Broder2000}}. MinHash builds a sketch for each set. A MinHash sketch consists of $k$ 32-bit registers.\\
\noindent $\bullet$ \textbf{HyperLogLog~\cite{FlajoletAOFA07}}. 
A HyperLogLog sketch consists of $k$ 5-bit registers,
and is originally designed for estimating a set's cardinality.
One can easily obtain a HyperLogLog sketch of $A\cup B$ by merging the HyperLogLog sketches of sets $A$ and $B$
and then use the sketch to estimate $|A \cup B|$.
Therefore, HyperLogLog can also be used to estimate $J_{A,B}$ by approximating $\frac{|A|+|B|-|A\cup B|}{|A\cup B|}$.\\
\noindent $\bullet$ \textbf{HyperMinHash~\cite{YuArxiv2017}}. A HyperMinHash sketch consists of $k$ $q$-bit registers and $k$ $r$-bit registers.
The first $k$ $q$-bit registers can be viewed as a HyperLogLog sketch.
To guarantee the performance for large sets (including up to $2^{32}$ elements), we set $q=5$.

\subsection{Metrics}
We evaluate both efficiency and effectiveness of our methods in comparison with the above baseline methods.
For efficiency, we evaluate the running time of all methods. 
Specially, we study the time for updating each set element and estimating set similarities, respectively.
The update time determines the maximum throughput that a method can handle,
and the estimation time determines the delay in querying the similarity of set-pairs.
For effectiveness, we evaluate the error of estimation $\hat{J}$ with respect to its true value $J$ using metrics: bias and root mean square error (RMSE), 
i.e., $\text{Bias}(\hat{J})=\mathbb{E}(\hat{J})-J$ and $\text{RMSE}(\hat{J})=\sqrt{\mathbb{E}((\hat{J} - J)^2)}$.
Our experimental results are empirically computed from $1,000$ independent runs by default.
We further evaluate our method on the detection of association rules,
and use \emph{precision} and \emph{recall} to evaluate the performance. 

\begin{figure}[t!]
	\centering
	\subfigure[$m=2^9$, $J_{A,B}=0.8$]{
		\includegraphics[width=0.22\textwidth]{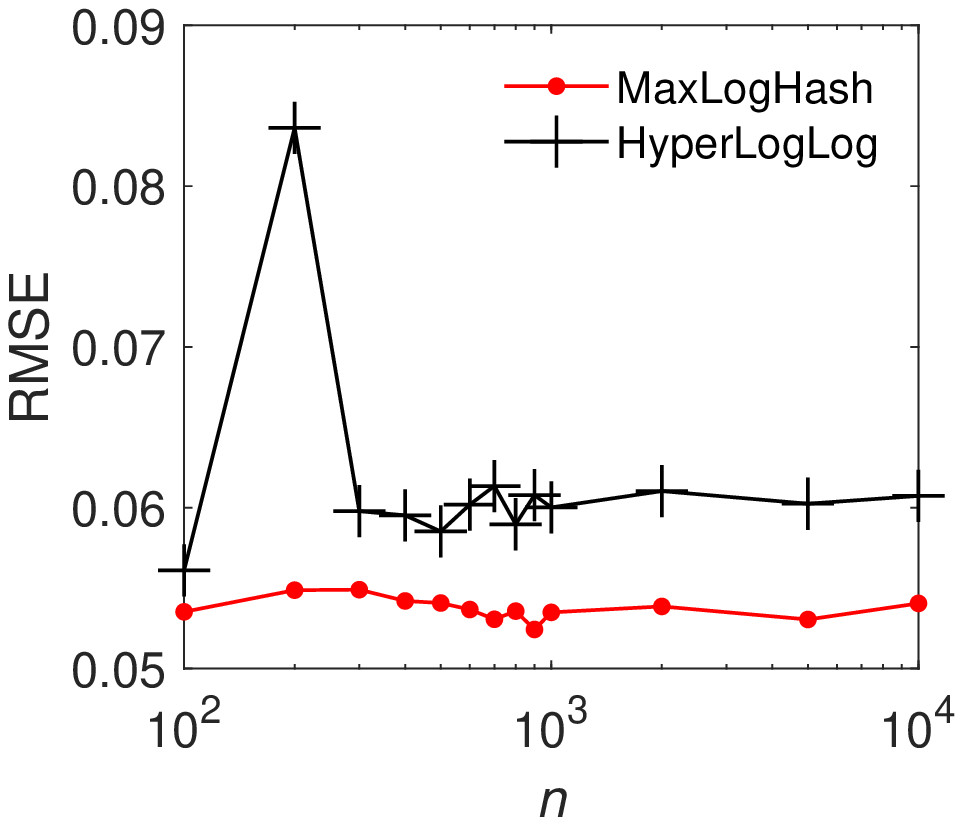}}
	\subfigure[$m=2^{9}$, $J_{A,B}=0.9$]{
		\includegraphics[width=0.22\textwidth]{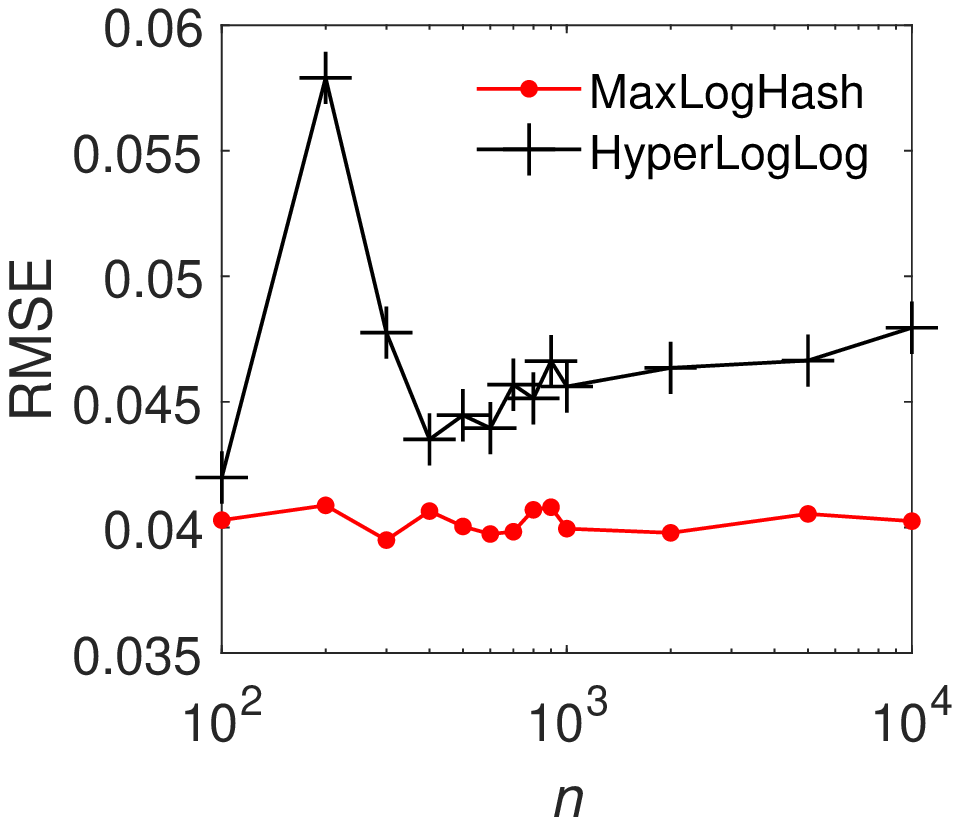}}
	\subfigure[$m=2^{10}$, $J_{A,B}=0.8$]{
		\includegraphics[width=0.22\textwidth]{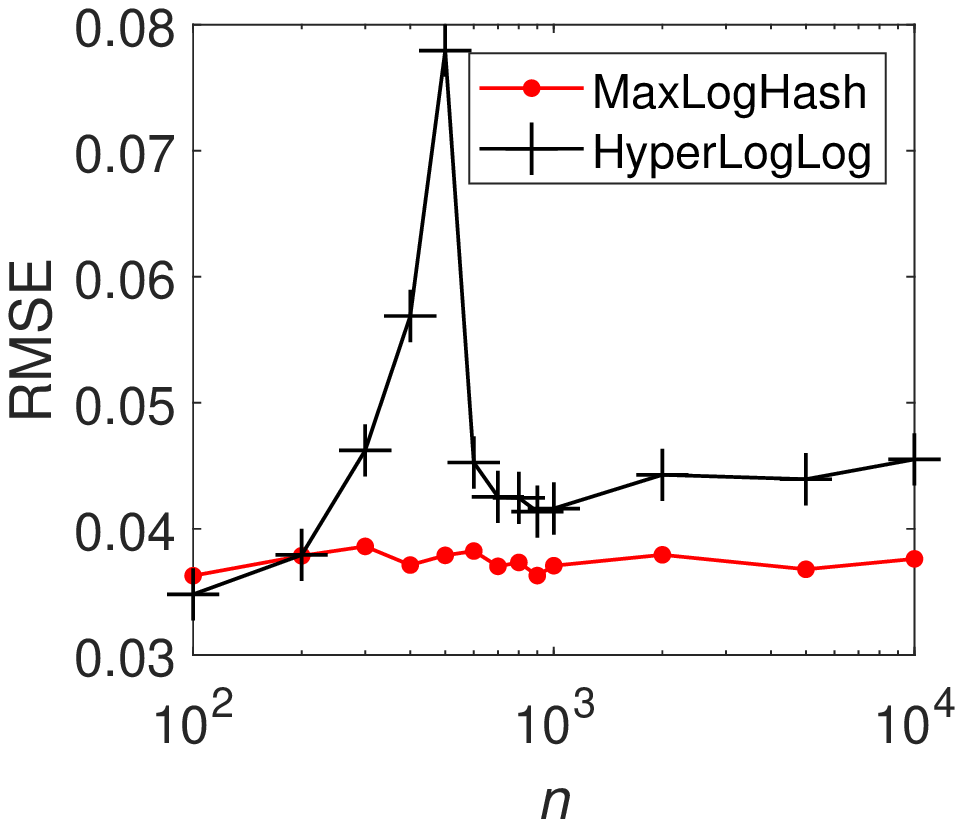}}
	\subfigure[$m=2^{10}$, $J_{A,B}=0.9$]{
		\includegraphics[width=0.22\textwidth]{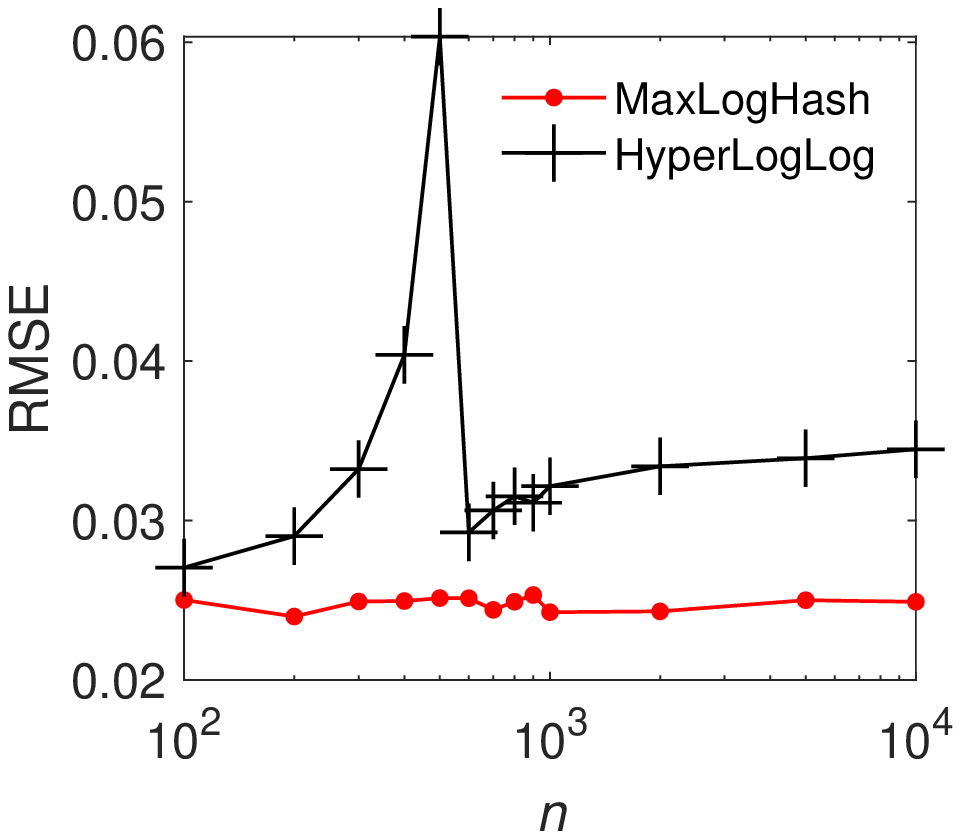}}
	\caption{Estimating error of our method MaxLogHash in comparison with HyperLogLog on synthetic set-pairs $A$ and $B$ with the same memory space $m$ bits, where $|A|=|B|=n$.}\label{fig:hllcard}
\end{figure}
\subsection{Accuracy of Similarity Estimation}
\textbf{MaxLogHash vs MinHash and HyperMinHash}.
From Figures~\ref{fig:synerror} (a)-(d), we see that our method MaxLogHash gives comparable results to MinHash and HyperMinHash with $r=4$.
Specially, the RMSEs of these three methods differ within $0.006$ and continually decrease as the similarity increases.
The RMSE of HyperMinHash with $r=1$ significantly increases as $J_{A,B}$ increases.
We observe that the large estimation error occurs because HyperMinHash exhibits a large estimation bias.
Figures~\ref{fig:synerror} (e)-(h) show the bias of our method MaxLogHash in comparison with MinHash and HyperMinHash.
We see that the empirical biases of MaxLogHash and MinHash are both very small and no systematic biases can be observed.
However, HyperMinHash with $r=1$ shows a significant bias and its bias increases as the similarity value increases.
To be more specific, its bias raises from $-0.06$ to $-0.089$ when  the similarity increases from $0.80$ to $0.99$.
One can increase $r$ to reduce the bias of HyperMinHash.
However, HyperMinHash with large $r$ desires more memory space.
For example, HyperMinHash with $r=4$ has comparable accuracy but requires $1.5$ times more memory space compared to our method MaxLogHash.
Compared with MinHash, MaxLogHash gives a $5.4$ times reduction in memory usage while achieves a similar estimation accuracy.
Later in Section~\ref{subsec:time}, we show that our method MaxLogHash has a computational cost similar to Minhash,
but is several orders of magnitude faster than HyperMinHash when estimating set similarities.

\noindent\textbf{MaxLogHash vs HyperLogLog}.
To make a fair comparison, we allocate the same amount of memory space, $m$ bits, to each of MaxLogHash and HyperLogLog.
As discussed in Section~\ref{sec:method}, the attractive property of our method MaxLogHash is its estimation error is almost independent from the cardinality of sets $A$ and $B$, 
which does not hold for HyperLogLog.
Figure~\ref{fig:hllcard} shows the RMSEs of MaxLogHash and HyperLogLog on sets of different sizes.
We see that the RMSE of our method MaxLogHash is almost a constant.
Figures~\ref{fig:hllcard} (a) and (b) show the performance of HyperLogLog suddenly degrades when  $m=2^{9}$ and the cardinalities of $A$ and $B$ are around $200$,
because HyperLogLog uses two different estimators for cardinalities within two different ranges respectively~\cite{FlajoletAOFA07}.
As a result, our method MaxLogHash decreases the RMSE of HyperLogLog by up to $36\%$.
As shown in Figures~\ref{fig:hllcard} (c) and (d), similarly, the RMSE of our method MaxLogHash is about 2.5 times smaller than HyperLogLog when $m=2^{10}$ and the cardinalities of $A$ and $B$ are around $500$.

\begin{figure}[t!]
	\centering
	\subfigure[balanced]{
		\includegraphics[width=0.22\textwidth]{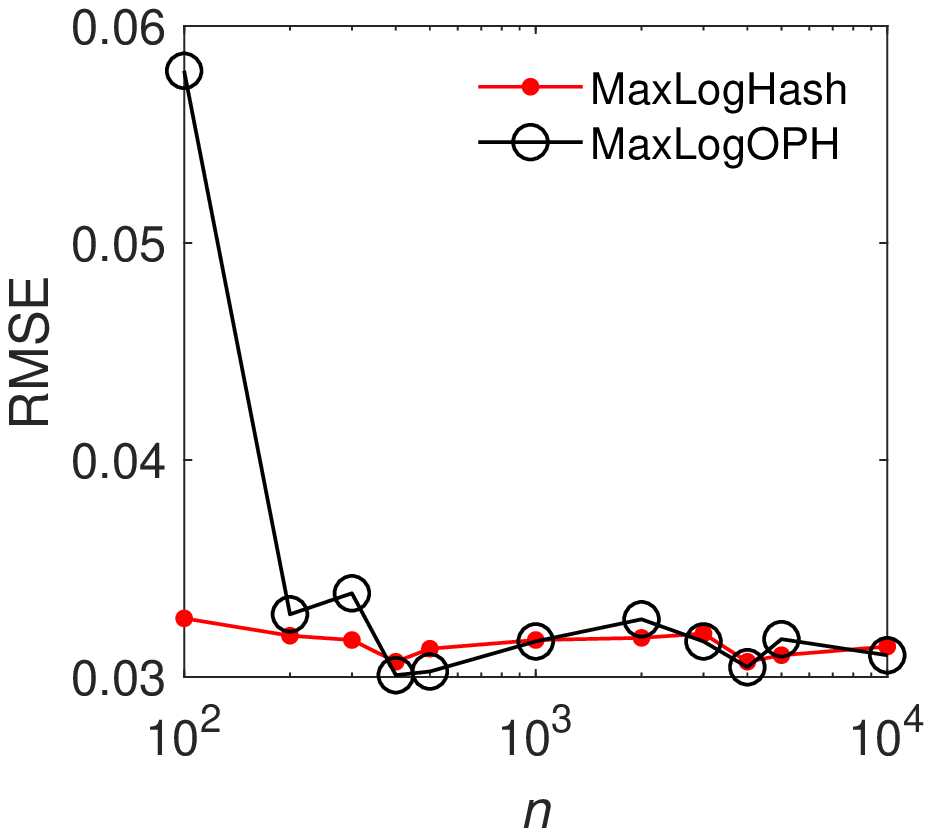}}
	\subfigure[unbalanced]{
		\includegraphics[width=0.22\textwidth]{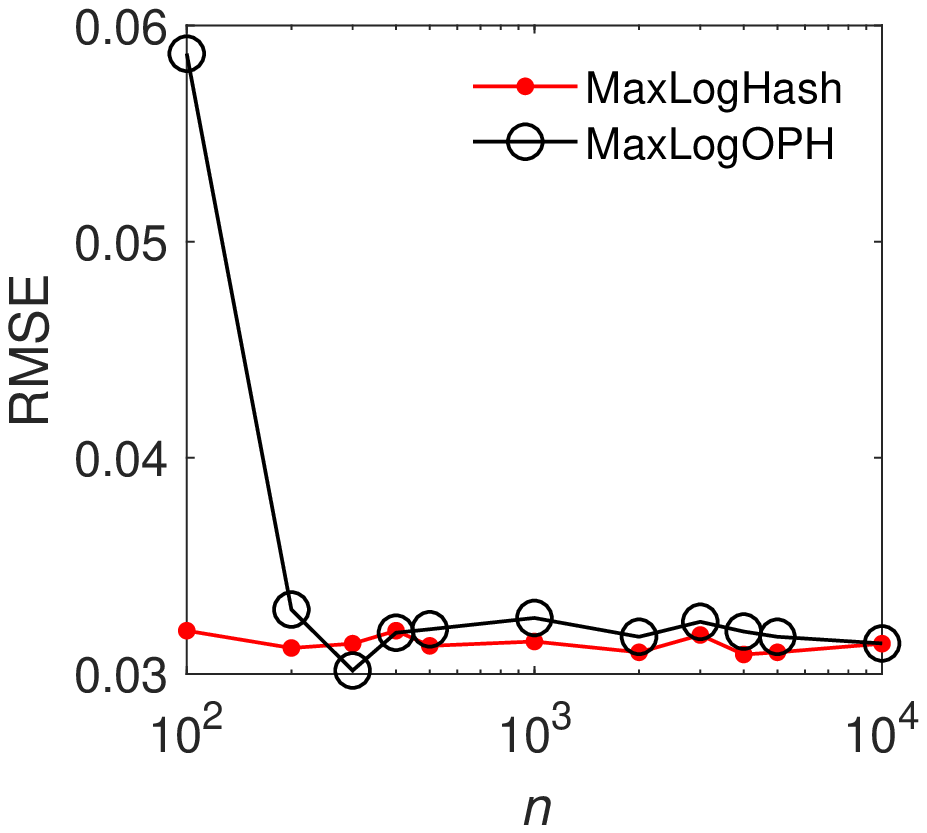}}
	\caption{Estimation error of our methods MaxLogHash and MaxLogOPH on both balanced and unbalanced synthetic data pairs $A$ and $B$ with the same number of registers, $k=128$, and  $J_{A,B}=0.9$. (a) $|A|=|B|=n$. (b) $|A|=J_{A,B}|B|=n$.}\label{fig:HashvsOPH}
\end{figure}

\noindent\textbf{MaxLogHash vs MaxLogOPH}. As discussed in Section~\ref{subsec:extension}, the estimation error of MaxLogOPH is comparable to MaxLogHash when $k$ is far smaller than the cardinalities of two sets of interest.
We compare MaxLogOPH with MaxLogHash on sets with increasing cardinalities to provide some insights.
As shown in Figure~\ref{fig:HashvsOPH}, MaxLogOPH exhibits relatively large estimation errors for small cardinalities.
When $k=128$ and the cardinality increases to $200$ (about $2k$), we see that MaxLogOPH achieves similar accuracy to MaxLogHash.
Later in Section~\ref{subsec:time}, MaxLogOPH significantly accelerates the speed of updating elements compared with MaxLogHash.

\subsection{Accuracy of Association Rule Learning}
\begin{figure*}[t]
	\centering
	\subfigure[(MUSHROOM) Precision, $J_0=0.8$]{
		\includegraphics[width=0.22\textwidth]{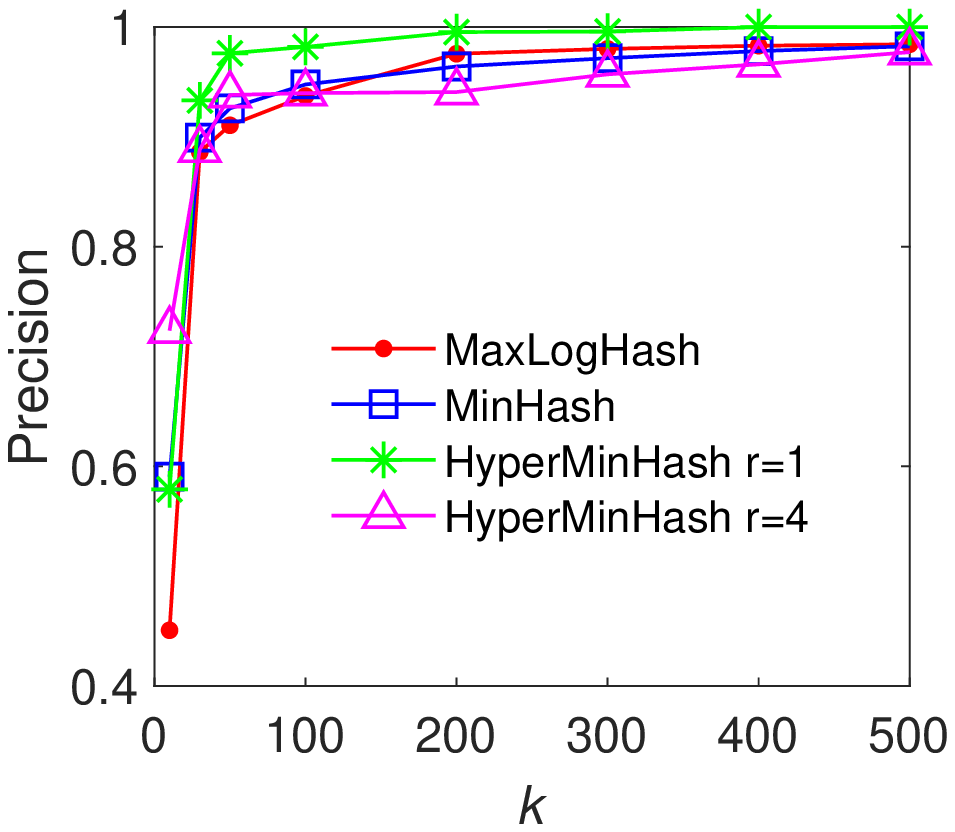}}
	\subfigure[(MUSHROOM) Recall, $J_0=0.8$]{
		\includegraphics[width=0.22\textwidth]{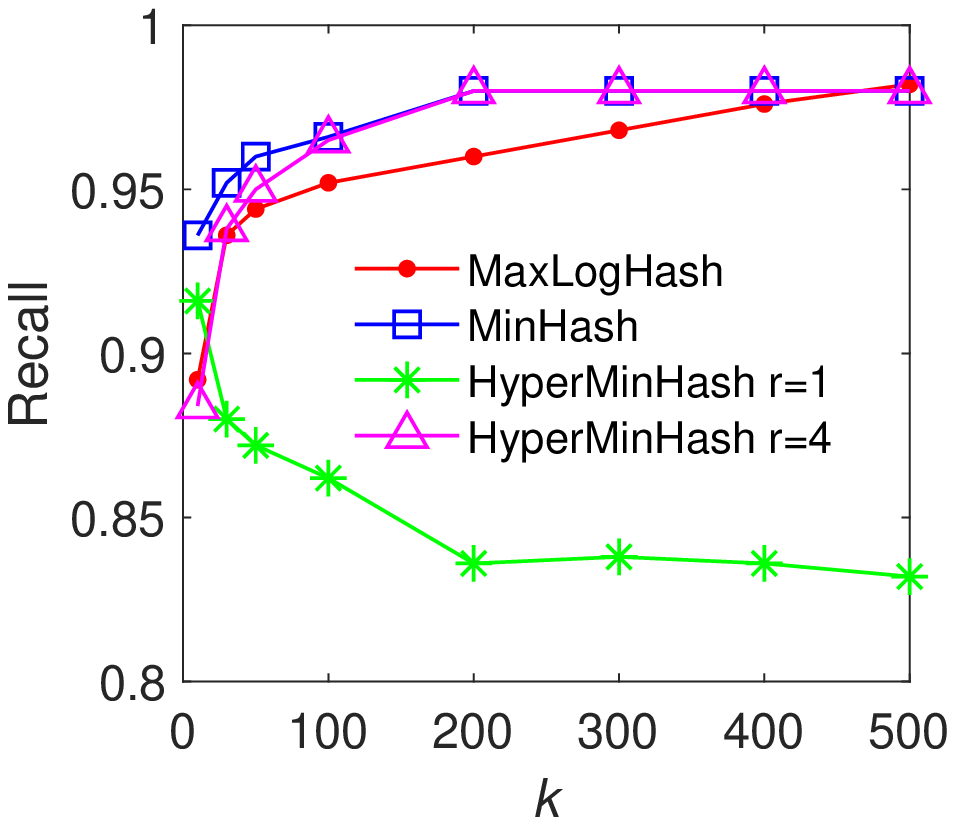}}
	\subfigure[(MUSHROOM) Precision, $J_0=0.9$]{
		\includegraphics[width=0.22\textwidth]{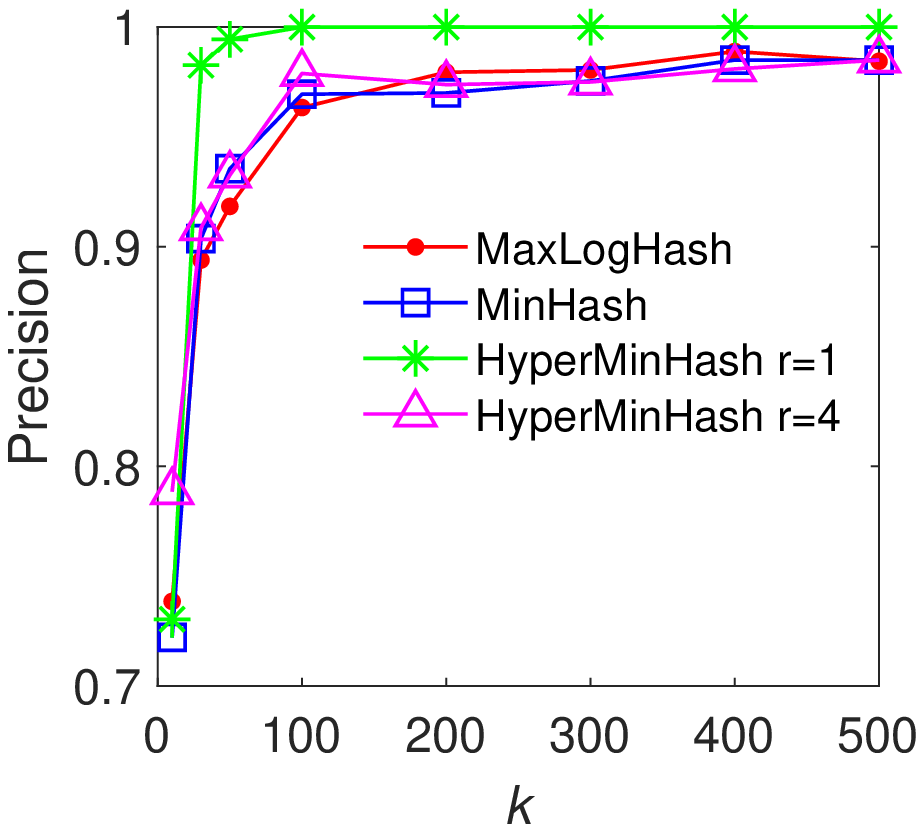}}
	\subfigure[(MUSHROOM) Recall, $J_0=0.9$]{
		\includegraphics[width=0.22\textwidth]{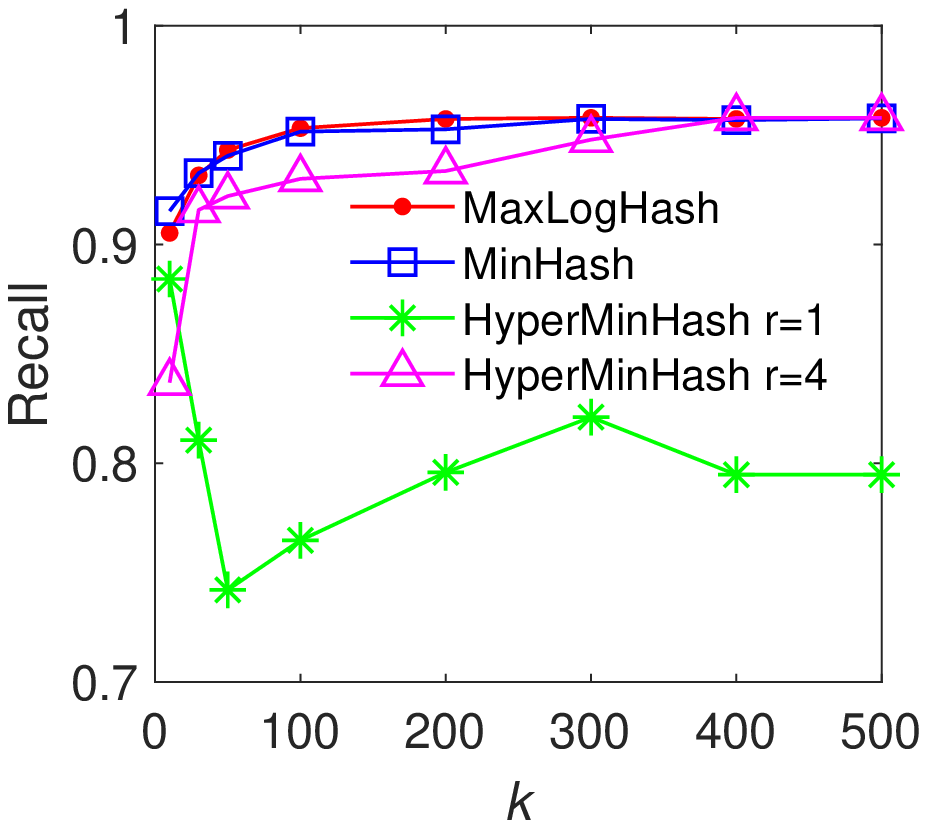}}
	\subfigure[(CONNECT) Precision, $J_0=0.8$]{
		\includegraphics[width=0.22\textwidth]{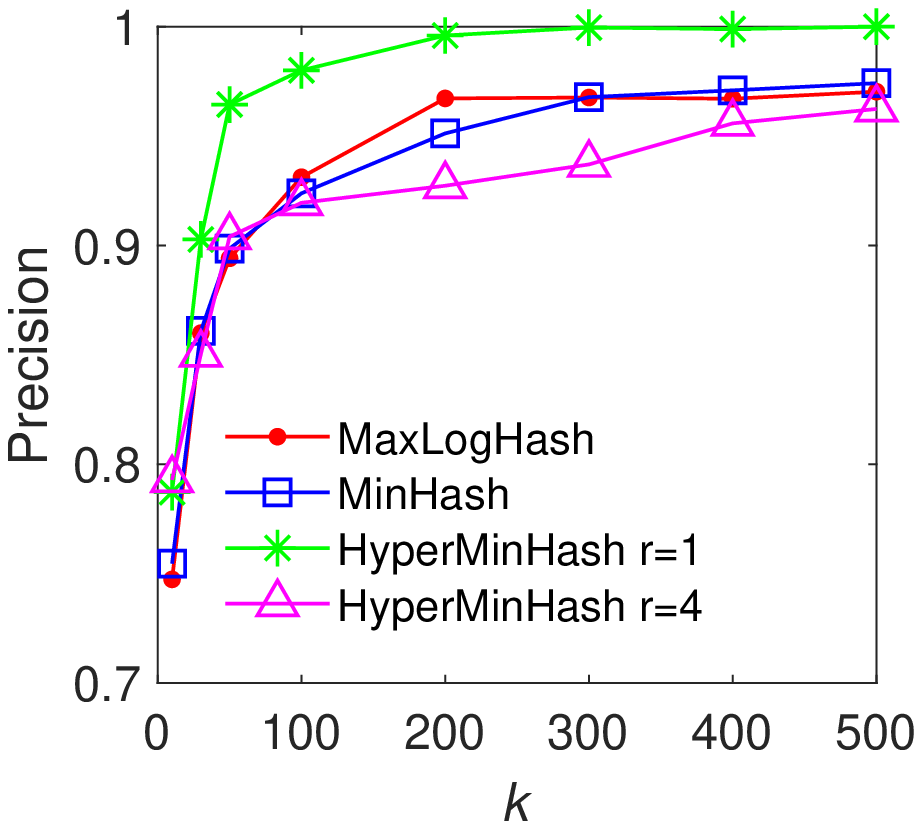}}
	\subfigure[(CONNECT) Recall, $J_0=0.8$]{
		\includegraphics[width=0.22\textwidth]{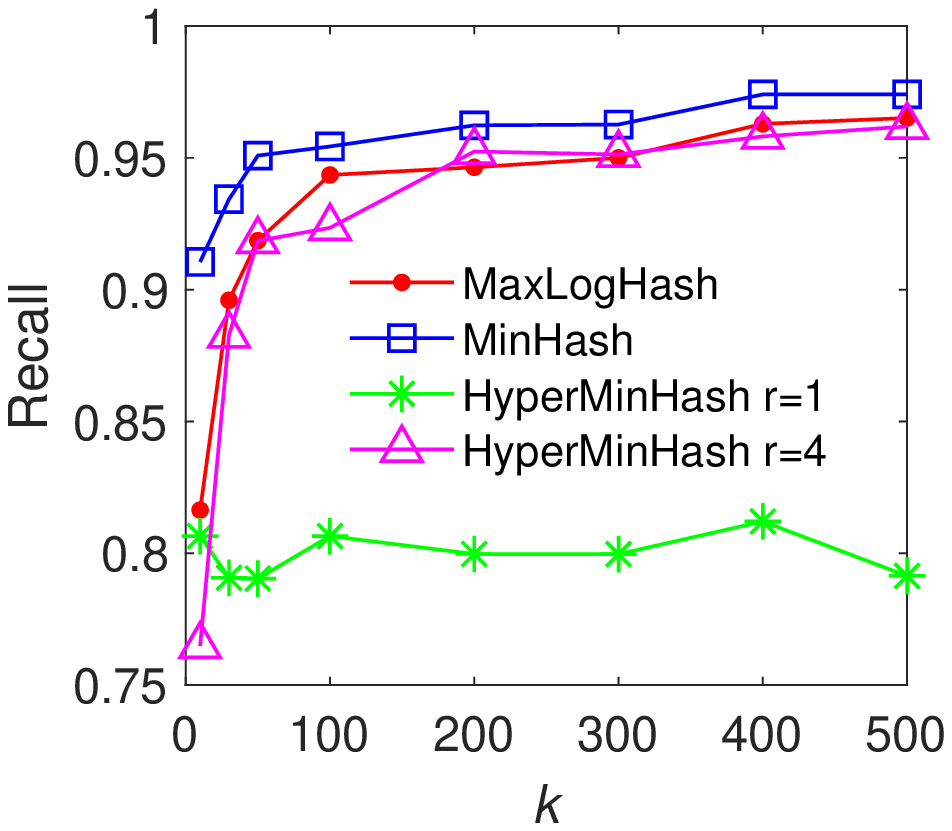}}
	\subfigure[(CONNECT) Precision, $J_0=0.9$]{
		\includegraphics[width=0.22\textwidth]{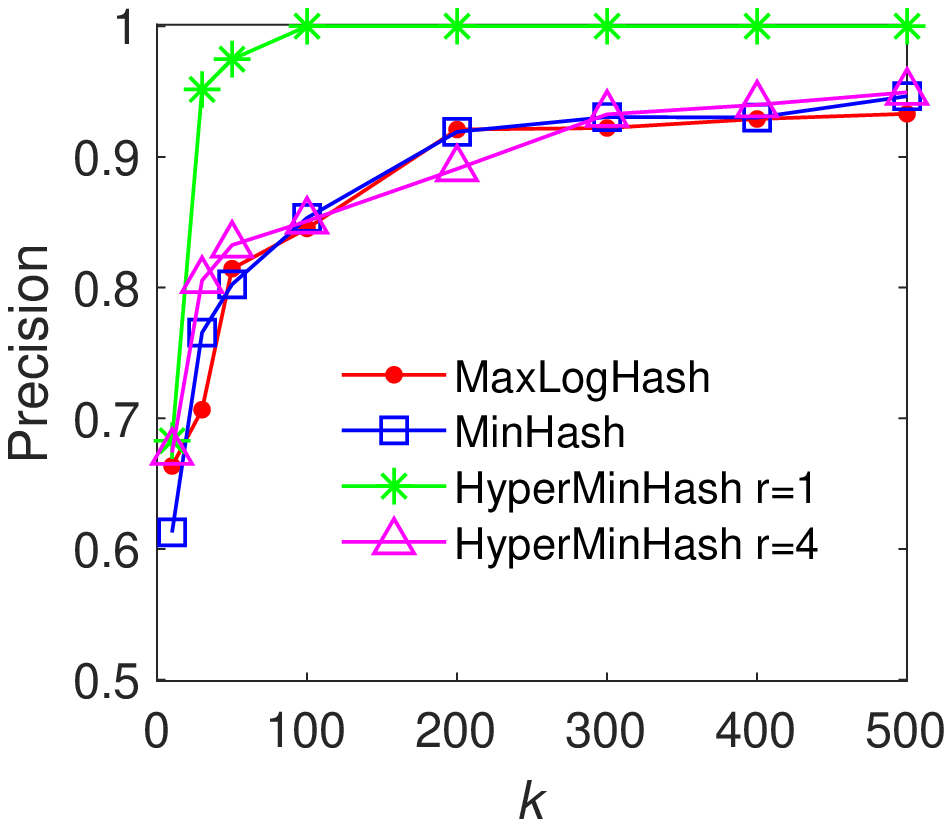}}
	\subfigure[(CONNECT) Recall, $J_0=0.9$]{
		\includegraphics[width=0.225\textwidth]{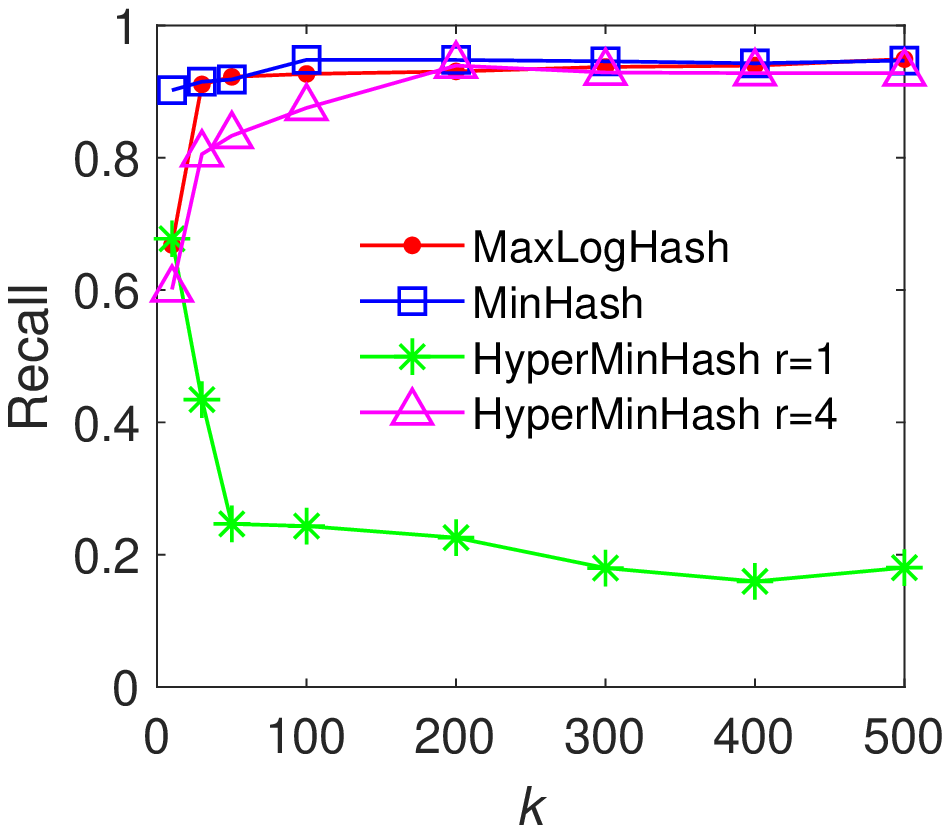}}
	\caption{Precision and recall of our method MaxLogHash in comparison with MinHash and HyperMinHash on datasets MUSHROOM and CONNECT.}\label{fig:precision}
\end{figure*}

In this experiment, we evaluate the performance of our method MaxLogHash, MinHash, and HyperMinHash on the detection of items (e.g., products) that almost always appear together in the same records (e.g., transactions).
We conduct the experiments on real-world datasets: MUSHROOM and CONNECT.
We first estimate all pairwise similarities among items' record-sets, and retrieve every pair of record-sets with similarity $J > J_0$. 
As discussed previously (results in Figure~\ref{fig:hllcard}), HyperLogLog is not robust, because it exhibits large estimation errors for sets of particular sizes.
Therefore, in what follows we compare our method MaxLogHash only with MinHash and HyperMinHash.
As shown in Figure~\ref{fig:precision}, MaxLogHash gives comparable precision and recall to MinHash and HyperMinHash with $r=4$.
We note that MaxLogHash gives up to $5.4$ and $2.4$ times reduction in memory usage in comparison with MinHash and HyperMinHash respectively.

\subsection{Efficiency}\label{subsec:time}
\begin{figure}[t]
	\centering
	\subfigure[update time]{
		\includegraphics[width=0.22\textwidth]{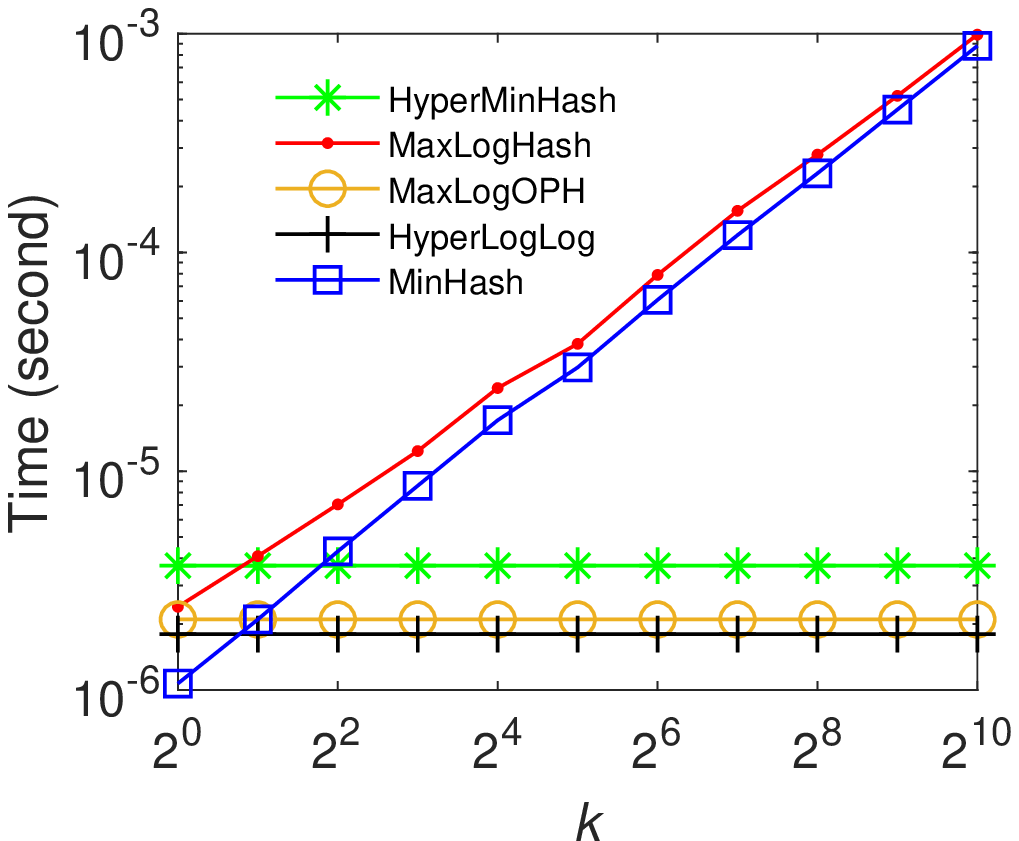}}
	\subfigure[estimation time]{
		\includegraphics[width=0.22\textwidth]{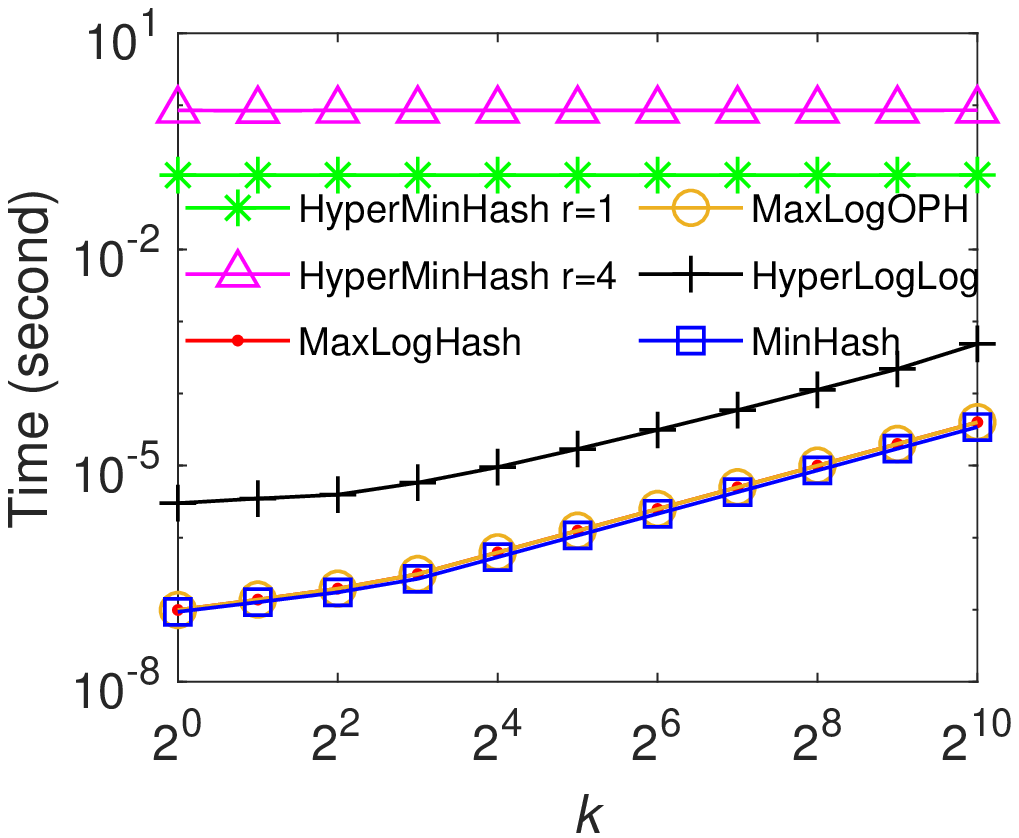}}
	\caption{Computational cost of our methods MaxLogHash and MaxLogOPH in comparison with MinHash, HyperLogLog and HyperMinHash.}\label{fig:efficiency}
\end{figure}

We further evaluate the efficiency of our method MaxLogHash and its extension MaxLogOPH in comparison with MinHash and HyperLogLog. 
Specially, we present the time for updating each coming element and computing Jaccard similarity, respectively.
We conduct experiments on synthetic balanced datasets.
We omit the similar results for real-world datasets and synthetic unbalanced datasets.
Figure~\ref{fig:efficiency} (a) shows that the update time of MaxLogOPH and HyperLogLog is almost a constant and our method outperforms other baselines.
The update time of HyperMinHash is almost irrelevant to its parameter $r$ and thus we only plot the curve for $r=1$.
Specially, MaxLogOPH is about $2$ and $420$ times faster than HyperMinHash and MinHash.
Figure~\ref{fig:efficiency} (b) shows that our methods MaxLogHash and MaxLogOPH have estimation time similar to MinHash,
while they are about $10$ times faster than HyperLogLog and 4 to 5 orders of magnitude faster than HyperMinHash.

\section{Related Work} \label{sec:related}
\noindent\textbf{Jaccard similarity estimation for static sets.} 
Broder et al.~\cite{Broder2000} proposed the first sketch method MinHash to compute the Jaccard similarity of sets,
which builds a sketch consisting of $k$ registers for each set.
To reduce the amount of memory space required for MinHash,
\cite{PingWWW2010,MitzenmacherWWW14} developed methods $b$-bit MinHash and Odd Sketch,
which are dozens of times more memory efficient than the original MinHash.
The basic idea behind $b$-bit MinHash and Odd Sketch is to use probabilistic methods such as sampling and bitmap sketching to build a compact digest for each set's MinHash sketch.
Recently, several methods~\cite{Linips2012,ShrivastavaUAI2014,ShrivastavaICML2014,ShrivastavaICML2017} were proposed to reduce the time complexity of processing each element in a set from $O(k)$ to $O(1)$.

\noindent\textbf{Weighted similarity estimation for static vectors.} 
SimHash (or, sign normal random projections)~\cite{CharikarSTOC2002} was developed for approximating angle similarity (i.e., cosine similarity) of weighted vectors.
CWS~\cite{Manasse2010,HaeuplerMT2014}, ICWS~\cite{IoffeICDM2010}, 0-bit CWS~\cite{LiKDD2015}, CCWS~\cite{WuICDM2016}, Weighted MinHash~\cite{ShrivastavaNIPS2016}, PCWS~\cite{WuWWW2017}, and BagMinHash~\cite{Ertl2018}  were developed for approximating generalized Jaccard similarity of weighted vectors\footnote{The Jaccard similarity between two positive real value vectors $\vec x = (x_1, x_2, \ldots, x_p)$ and $\vec y = (y_1, y_2, \ldots, y_p)$  is defined as $J(\vec x, \vec y) = \frac{\sum_{1\le j\le p}\min (x_j, y_j)}{\sum_{1\le j\le p}\max (x_j, y_j)}$.},
and Datar et al.~\cite{DatarSOCG2004} developed an LSH method using $p$-stable distribution for estimating $l_p$ distance for weighted vectors,
where $0<p\le 2$.
Campagna and Pagh~\cite{CampagnaKAIS2012} developed a biased sampling method for estimating a variety of set similarity measures beyond Jaccard similarity.

\noindent\textbf{Similarity estimation for data streams.}
The above weighted similarity estimation methods fail to deal with streaming weighted vectors, whereas elements in vectors come in a stream fashion.
To solve this problem, Kutzkov et al.~\cite{KutzkovCIKM2015}  extended AMS sketch~\cite{AlonSTOC1996} for the estimation of cosine similarity and Pearson correlation in streaming weighted vectors.
Yang et al.~\cite{YangICDM2017} developed a streaming method HistoSketch for approximating Jaccard similarity with concept drift.
Set intersection cardinality (i.e., the number of common elements in two sets) is also a popular metric for evaluating the similarity in sets.
A variety of sketch methods such as
LPC~\cite{Whang1990}, FM~\cite{Flajolet1985},
LogLog~\cite{Durand2003}, HyperLogLog~\cite{FlajoletAOFA07}, HLL-TailCut+~\cite{XiaoZC17},
and MinCount~\cite{GiroireDAMNew2009} were proposed to estimate the stream cardinality (i.e., the number of distinct elements in the stream),
and can be easily extended to estimate $|A\cup B|$ by merging the sketches of sets $A$ and $B$.
Then, one can approximate $|A\cap B|$ because  $|A\cap B|= |A| + |B| - |A\cup B|$.
To further improve the estimation accuracy, 
Cohen et al.~\cite{CohenKDD2017} developed a method combining MinHash and HyperLogLog to estimate set intersection cardinalities.
Our experiments reveal that these sketch methods have large errors when first estimating $|A\cap B|$ and $|A\cup B|$, and then approximating the Jaccard similarity $J_{A,B}$.
As mentioned in Section~\ref{sec:preliminaries}, MinHash can be easily extended to handle streaming sets,
but its two compressed versions, $b$-bit MinHash and Odd Sketch fail to handle data streams.
To solve this problem,
Yu and Weber~\cite{YuArxiv2017} developed a method, HyperMinHash, which can be viewed as a joint of HyperLogLog and $b$-bit MinHash.
HyperMinHash consists of $k$ registers,
whereas each register has two parts, an FM sketch and a $b$-bit string.
The $b$-bit string is computed based on the fingerprints (i.e., hash values) of set elements that map to the register.
HyperMinhash first estimates $|A\cup B|$ and then infers the Jaccard similarity of sets $A$ and $B$ from the number of collisions of $b$-bit strings
given $|A\cup B|$.
Our experiments demonstrates that HyperMinHash exhibits a large bias for high similarities
and it is several orders of magnitude slower than our methods when estimating the similarity.

\section{Conclusions and Future Work} \label{sec:conclusions}
We develop a memory efficient sketch method MaxLogHash to estimate the similarity of two sets given in a streaming fashion.
We provide a simple yet accurate estimator for Jaccard similarity, and derive exact formulas for the estimator's bias and variance.
Experimental results demonstrate that MaxLogHash can reduce around 5 times the amount of memory required for MinHash with the same desired accuracy and computational cost.
Compared with our method MaxLogHash,
the state-of-the-art method HyperMinHash exhibits a larger estimation bias and its estimation time is 4 to 5 orders of magnitude larger.
Although HyperLogLog can be extended to estimate Jaccard similarity,
its estimation error (resp. estimation time) is about 2.5 times (resp. 10 times) larger than our methods.
In the future, we plan to extend MaxLogHash to weighted streaming vectors and fully dynamic streaming sets that include both set element insertions and deletions.

\section*{Acknowledgment}
The research presented in this paper is supported in part by National Key R\&D Program of China (2018YFC0830500), National Natural Science Foundation of China (U1736205, 61603290), Shenzhen Basic Research Grant (JCYJ20170816100819428), Natural Science Basic Research Plan in Shaanxi Province of China (2019JM-159).
The work of John C.S. Lui is supported in part by the GRF R4032-18.

\balance
\bibliographystyle{unsrt}
\bibliography{COPH,randpe,ctstream,simcar,albitmap}

\end{document}